\newcommand{\ud}{\rm d}
\newcommand{\un}{~\mathrm}

\documentclass{iopart}

\usepackage[T1]{fontenc}% pour utiliser les caractères français 
\usepackage[pdftex]{graphicx}
\usepackage{dcolumn}% Align table columns on decimal point 
\usepackage{bm}% bold math 
\usepackage[colorlinks=true]{hyperref}
\usepackage{lscape}

\begin{document}
\review{Intermittency and roughening in the failure of brittle heterogeneous materials}

\author{Daniel Bonamy}
\address{CEA, IRAMIS, SPCSI, Grp. Complex Systems $\&$ Fracture, F-91191 Gif sur Yvette, France}

\begin{abstract}
Stress enhancement in the vicinity of brittle cracks makes the macro-scale failure properties extremely sensitive to the micro-scale material disorder. Therefore: (i) Fracturing systems often display a jerky dynamics, so-called crackling noise, with seemingly random sudden energy release spanning over a broad range of scales, reminiscent of earthquakes; (ii) Fracture surfaces exhibit roughness at scales much larger than that of material micro-structure. Here, I provide a critical review of experiments and simulations performed in this context, highlighting the existence of universal scaling features, independent of both the material and the loading conditions, reminiscent of critical phenomena. I finally discuss recent stochastic descriptions of crack growth in brittle disordered media that seem to capture qualitatively - and sometimes quantitatively - these scaling features.
\end{abstract}
 
\submitto{\JPD}
\maketitle
\tableofcontents
\newpage
% body of paper here

\section{Introduction}\label{Sec:1}

Understanding how materials break is of fundamental and technological interest. As such, it has attracted much attention from engineers and physicists. Since the pioneer work of Griffith \cite{Griffith20_ptrs}, a coherent theoretical framework, the Linear Elastic Fracture Mechanics (LEFM) was developed. It states that, in an elastic material, crack initiation occurs when the mechanical energy released by crack advance is sufficient to balance that needed to create new surfaces \cite{Griffith20_ptrs}. Then, once the crack starts to grow, inertial effects should be included in the energy balance (see e.g. \cite{Freund90_book}). This continuum theory has proven to be extremely powerful to describe ideal {\em homogeneous} media, but fails to capture some features observed in heterogeneous ones. 

While the mechanical energy released during crack growth is well determined by continuum theory, the dissipation processes occur within a tiny zone at the crack tip. This separation in length-scales, intrinsic to fracture mechanics, makes the failure properties observed at the continuum length-scale extremely sensitive to the microstructural disorder of the material. Consequences include important fluctuations in the strength exhibited by different samples of the same material \cite{Weibull39_book}, size effects in the strength \cite{Bazant98_book,Alava09_issue}, crackling dynamics with seemingly random violent events of energy release prior and during the failure \cite{Gutenberg54_book} and rough crack paths \cite{Hull99_book}, among others. 

I will focus in this review on two of these aspects, namely the crackling dynamics evidenced in brittle failure and the morphology of fracture surfaces. Experimental observations performed in this context report the existence of scale invariant laws \cite{Omori94_rieic,Gutenberg54_book,Mandelbrot84_nature,Bouchaud97_jpcm}. This has attracted much attention from the statistical physics community over the last 25 years \cite{Herrmann90_book,Chakrabarti97_book,Alava06_ap}. In this respect, lattice models such as Random Fuse Models (RFM) \cite{deArcangelis85_jpl} were shown to capture qualitatively these scaling features in minimal models,  keeping only the two key ingredients responsible for the complexity in material breakdown: The material local disorder and the long-range elastic load redistribution after a local failure event. More recently, it was proposed to extend LEFM descriptions to heterogeneous brittle materials by taking into account the microstructural disorder through a stochastic term \cite{Gao89_jam,Schmittbuhl95_prl}. As a result, the onset of crack propagation is analogue to a dynamic phase transition between a stable phase, where the crack remains pinned by material heterogeneities, and a moving phase, where the mechanical energy available at crack tip is sufficient to make the front propagate. As such, it exhibits universal - and predictable - scaling laws that reproduce fairly well the intermittency statistics observed in crack dynamics \cite{Bonamy08_prl} and the morphological scaling features exhibited by fracture surfaces \cite{Bonamy06_prl}. 
  
This review is organized as follow. First, I provide a brief introduction to standard LEFM theory in section \ref{Sec:2} and discuss some of its predictions in term of fracture dynamics and crack roughness. Then, I review in section \ref{Sec:3} the experiments and fields observations that have allowed to characterize the intermittent dynamics observed in many fracturing systems, from earthquakes to laboratory fracture experiments. Section \ref{Sec:4} is devoted to a critical overview of the morphological scaling features evidenced experimentally in various materials. RFM models and their predictions in term of crackling dynamics and crack roughness are discussed in section \ref{Sec:5}. Section \ref{Sec:6} discusses the stochastic descriptions of crack growth and highlights their predictions in term of scale-free distributions, scaling laws and scale-invariant morphology for fracture in perfectly brittle disordered materials. Finally, section \ref{Sec:7} outlines the current challenges and possible perspectives.  

\section{Continuum theory of crack growth}\label{Sec:2}

\subsection{Stress concentration}\label{Sec:2.1}

\begin{figure}
\begin{center}
\includegraphics[width=0.95\textwidth]{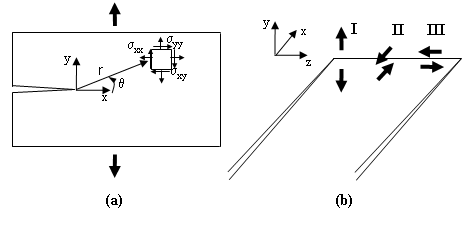}
\caption{(a) Sketch and notations describing the stress field in the vicinity of a slit crack tip in a two-dimensional medium under tension. (b) The three modes of fracture: Tensile mode (Mode I), shearing mode (Mode II) and tearing mode (mode III).}
\label{Sec:2:fig1}
\end{center}
\end{figure}

Crack propagation is the basic mechanism leading to the failure of brittle materials. The crack motion in a brittle homogeneous material is classically analyzed using methods from elastodynamics. Let us first consider the case of a straight running crack embedded in a linear elastic material under tension. As first noted by Orowan \cite{Orowan55_wrs} and Irwin \cite{Irwin57_jam}, the stress field $\sigma_{ij}$ is singular in the vicinity of the crack tip and can be written as (see e.g. \cite{Freund90_book}):

\begin{equation}
\sigma_{ij}(r,\theta)=\frac{K_I}{\sqrt{2 \pi r}} F_{ij}(\theta,v)+T G_{ij}(\theta,v)+O(\sqrt{r}),
\label{Sec:2:equ1}
\end{equation}

\noindent where $v$ is the crack speed, $(r,\theta)$ are the polar coordinates in the frame $(\vec{e}_x,\vec{e}_y)$ centered at the crack tip (see Figure \ref{Sec:2:fig1}a), and $F_{ij}(\theta,v)$ and $G_{ij}(\theta,v)$ are some non-dimensional universal functions indicating the angular variation of the stress field, and its variation with $v$. The two prefactors $K_I$ and $T$, called {\em Stress Intensity Factor} and {\em $T$-stress} respectively, only depend on the applied loading and the specimen geometry. They determine the intensity of the singular and non-singular parts of the local field. 

This can be generalized to any loading conditions. It is convenient to decompose them into the sum of three modes (Figure \ref{Sec:2:fig1}b):
\begin{itemize}
\item Mode $I$ (tensile mode) corresponds to normal separation of the crack under tensile stresses.
\item Mode $II$ (shearing mode) corresponds to a shear parallel to the direction of crack propagation.
\item Mode $III$ (tearing mode) corresponds to a shear parallel to the crack front. 
\end{itemize}
\noindent In the vicinity of the crack tip, the stress field is then written as a sum of three terms the shape of which is given by the right hand term of equation \ref{Sec:2:equ1} with prefactors $\{K_I,T_I\}$, $\{K_{II},T_{II}\}$ and $\{K_{III},T_{III}\}$ associated to the tensile, shearing and tearing modes, respectively. As we shall see in section \ref{Sec:2.3}, a crack propagating in an isotropic solid chooses its orientation so that it makes shear vanish at its tip. As a consequence, in most of the crack problems discussed thereafter, we will consider solids loaded dominantly in tension where the mode II and III perturbations will be subordinated to that in mode I. Two exceptions will be discussed later in this review, namely earthquakes problems and rocks broken under compression. In both cases shear and/or tear fracture modes are dominant.

\subsection{Crack propagation: Griffith theory}\label{Sec:2.2}

Because of the stress singularity, the mechanical energy released as fracture occurs is entirely dissipated within a small zone at the crack tip, referred to as the {\em Fracture Process Zone} (FPZ). In nominally brittle materials, the absence of an outer plastic zone can be assumed. According to Griffith's theory \cite{Griffith20_ptrs}, the onset of fracture is then reached when the amount of elastic energy released by the solid as the crack propagates by a unit length is equal to the {\em fracture energy} $\Gamma$ defined as the energy dissipated in the FPZ to create two new fracture surfaces of unit length. The form taken by the stress field at the crack tip (Equation \ref{Sec:2:equ1}) allows relating the mechanical energy release $G$ at the onset of crack propagation ($v=0$) to $K_I$:

\begin{equation}
G = \frac{K_I^2}{E},
\label{Sec:2:equ2}
\end{equation}

\noindent where $E$ is the Young modulus of the material. The Griffith criterion for fracture onset then reads:

 \begin{equation}
G \geq \Gamma \quad \mathrm{or}\,\,\mathrm{equivalently} \quad K_I \geq K_c \quad \mathrm{with} \quad K_c=\sqrt{E\Gamma},
\label{Sec:2:equ3}
\end{equation}

\noindent where the so-called {\em toughness} $K_c$ is, like $\Gamma$, a material property to be determined experimentally. Then, once the crack starts to propagate, its motion is governed by the balance between the mechanical energy flux into the FPZ and the dissipation rate $\Gamma v$. From the form taken by the stress field at the crack tip (Equation \ref{Sec:2:equ1}), the energy flux can be related to $K_I$ and $v$, which yields (see e.g. \cite{Freund90_book}):

 \begin{equation}
A(v)\frac{K_I^2}{E}=\Gamma \quad \mathrm{with} \quad A(v)\simeq \left( 1-\frac{v}{c_R}\right),
\label{Sec:2:equ4}
\end{equation}

\noindent where $c_R$ refers to the Rayleigh wave speed in the material. The dynamic fracture regime reached when $v$ is on the order of $c_R$ is beyond the scope of this review. An interested reader can refer to the reviews of Ravi-Chandar (1998) \cite{Ravichandar98_ijf} and Fineberg and Marder (1999)\cite{Fineberg99_pr} for a detailed presentation of this regime. In the sequel, I will focus on the slow crack growth regime where $v \ll c_R$. Then, the equation of motion can be rewritten as:

 \begin{equation}
\frac{1}{\mu} v \simeq K_I - K_{c} \quad \mathrm{or}\,\mathrm{equivalently} \quad \frac{1}{\mu'} v \simeq G - \Gamma,
\label{Sec:2:equ5}
\end{equation}
  
\noindent where the effective mobilities $\mu$ and $\mu'$ are given by $\mu=2c_R/K_c$ and $\mu'=c_R/\Gamma$, respectively. 

\subsection{Crack path: Principle of Local Symmetry}\label{Sec:2.3}

Finally, to complete the continuum theory of crack growth in ideal homogeneous brittle materials, one has to add a path criterion. This is provided by the {\em Principle of Local Symmetry} (PLS) of Goldstein and Salganik (1974) \cite{Goldstein74_ijf} which states that a moving crack progresses along a direction so as to remain in pure tension. In 2D systems, the crack is loaded by a combination of mode I and II only, and the PLS implies that the direction of crack propagation is chosen so that $K_{II}=0$. In 3D systems, the crack load can also contain a mode III component. Then, to cancel $K_{III}$ and to propagate in pure mode I, the crack front would have to twist abruptly around the direction of propagation, which would yield unphysical discontinuities in the crack path. In this situation, the front is commonly observed to split into many pieces and to form `lances' \cite{Sommer69_efm} the origin of which remains presently highly debated (see e.g. \cite{Lazarus01a_jmps,Lazarus01b_jmps} for recent works on this topic). 
 
\subsection{Continuum mechanics predictions}\label{Sec:2.4}

There are numerous consequences of the {\em Linear Elastic Fracture Mechanics} (LEFM) summarized above. Here I will focus on two of them: (i) The equation of motion (Equation \ref{Sec:2:equ5}) predicts {\em regular and continuous} dynamics for crack propagation; (ii) The PLS predicts {\em smooth surfaces} at continuum scales, i.e. at scales over which the mechanical properties of the considered brittle material are homogeneous. As one will see in the two next sections, there are numerous experimental observations that have been accumulated over the past three decades that contradict both of these predictions.

\section{Crackling dynamics in fracture}\label{Sec:3}

\subsection{Earthquake statistics}\label{Sec:3.1}

\begin{figure}
\begin{center}
\includegraphics[width=0.49\textwidth]{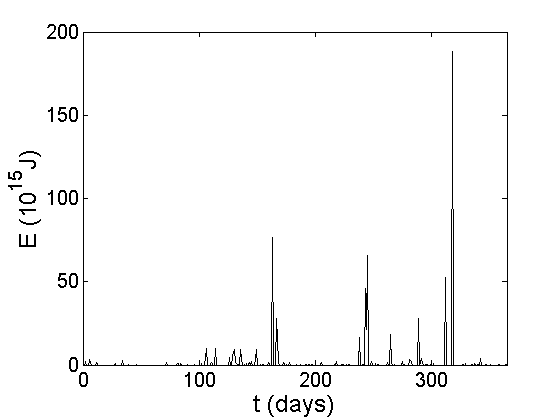}
\includegraphics[width=0.49\textwidth]{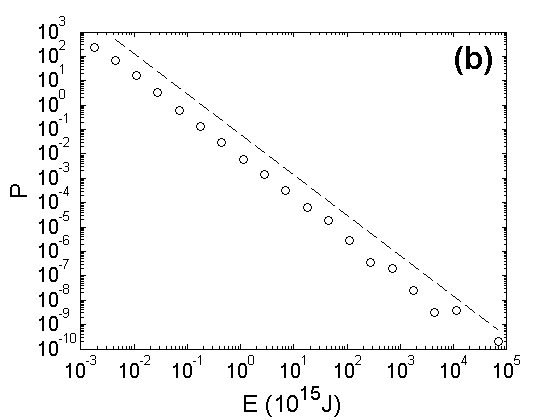}
\includegraphics[width=0.49\textwidth]{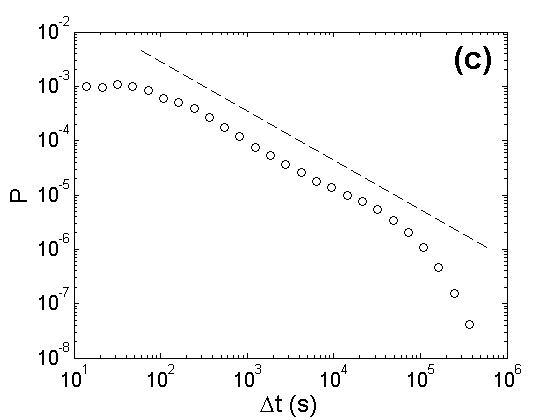}
\includegraphics[width=0.49\textwidth]{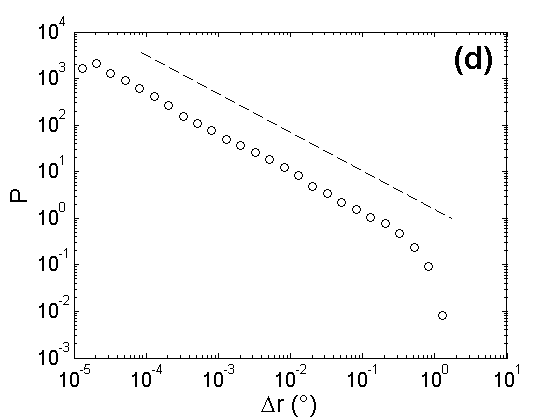}
\caption{(a) Energy radiated by earthquakes of magnitude $M \geq 3$  for each day of the year 2005, in a region of California spanning $32^\circ\mathrm{N}-37^\circ\mathrm{N}$ latitude and $114^\circ\mathrm{W}-122^\circ\mathrm{W}$ longitude. (b) Distribution in energy of for earthquakes having occurred in this region from 1987 to present. The axes are logarithmic. The dash line is the power-law given by equation \ref{Sec:3:equ1} with $\beta=1.7$. (c) Distribution of waiting time $\Delta t$ between two successive earthquakes of magnitude larger than $2.5$ within a zone of size $L=1^\circ$ in this region from 1987 to present. The axes are logarithmic. The dash line is the Omori's law (Equation \ref{Sec:3:equ2}) with $\alpha=0.9$. (c) Distribution of jumps size $\Delta r$ between two successive earthquakes of magnitude larger than $2.5$ within a zone of size $L=1^\circ$ in this region from 1987 to present. The axes are logarithmic. The dash line is the law of Davidsen and Paczuski (Equation \ref{Sec:3:equ3}) with $\delta=0.7$. Data were taken from seismic catalogs available at Southern California Earthquake Data Center (SCEDC) \cite{SCEDC}.}
\label{Sec:3:fig1}
\end{center}
\end{figure}

Contrary to that predicted by LEFM, crack growth in heterogeneous brittle materials often displays a jerky dynamics, with seemingly random discrete jumps of a variety of sizes. The most obvious signature of this crackling dynamics can be found in the seismic activity of faults: Earth responds to the slow shear strains imposed by the continental drift through a series of violent impulses, earthquakes, spanning over many orders of magnitude, from barely noticeable to catastrophic ones (Figure \ref{Sec:3:fig1}a). The distribution of radiated seismic energy presents the particularity to form a power-law (Figure \ref{Sec:3:fig1}b) with no characteristic scale:

\begin{equation}
P(E) \propto E^{-\beta},
\label{Sec:3:equ1}
\end{equation}

\noindent where the exponent $\beta \simeq 1.7$ was proven to be very robust over different regions \cite{Utsu99_pag,Benzion08_rg}.

Earthquake sizes are most commonly quantified by their magnitude $M$. The first (and most popular) magnitude scale is the one introduced by Richter (1935) \cite{Richter35_bssa} that relates $M$ to the logarithm of the maximum amplitude measured on a Wood-Anderson torsion seismometer at a given distance from the earthquake epicenter. In this respect, $M$ is relatively easy to measure and it provides information that is directly useful for various engineering applications. Magnitude and radiated seismic energy of earthquakes are usually related via the Gutenberg-Richter empirical relation \cite{Gutenberg56_bssa}:

\begin{equation}
\log_{10} E = 1.5 M +11.8,
\label{Sec:3:equ1bis}
\end{equation}

\noindent where $E$ is expressed in Joule. While this relation is considered to be reasonably accurate for earthquakes of small and intermediate sizes, its validity is questioned for great earthquakes \cite{Kanomori77_jgr}. Using this energy-magnitude relation in equation \ref{Sec:3:equ1} leads to the Gutenberg-Richter frequency-magnitude statistics:

\begin{equation}
p(M) \propto \exp(- b M), 
\label{Sec:3:equ1ter}
\end{equation}

\noindent where $b$ and $\beta$ are related through: $\beta=b/1.5+1$.

The $b$ measurement has been one of the most frequently discussed topics in seismicity studies (see e.g. \cite{Utsu99_pag} for a recent review). Observed $b$ values of regional seismicity typically fall in the range $0.7-1.3$ and usually take a value around $1$ \cite{Utsu99_pag}. Its universality (or not) remains an open question \cite{Utsu99_pag}, made particularly tough since $b$ measurement can be affected by various factors as e.g. the type of magnitude scale chosen, the completeness and the magnitude range of the collected data, to name a few.

The radiated seismic energy is also found to scale with the rupture area $S$ of earthquakes as \cite{Abe75_jpe,Kanomori75_bssa}:

\begin{equation}
E \propto S^{1.5}. 
\label{Sec:3:equ1quad}
\end{equation}

\noindent Using this relation in equation \ref{Sec:3:equ1} yields the following distribution for rupture area:

\begin{equation}
p(S) \propto S^{-\tau}, 
\label{Sec:3:equ1cinq}
\end{equation}

\noindent with $\tau = 1.5 \beta - 0.5 \simeq 2$.

The time occurrence of earthquakes exhibits also scale-free features expressed by Omori's law \cite{Omori94_rieic}, which states that a main earthquake is followed by a sequence of aftershocks the frequency of which decays with time as $t^{-\alpha}$, with $\alpha \simeq 1$. More recently, driven by the availability of complete seismic catalogs (see e.g. \cite{SCEDC}) gathering the occurrence time, the magnitude, and the 3D location of  earthquakes, many studies \cite{Bak02_prl,Kagan02_bssa,Ziv03_bssa,Corral04_prl,Davidsen05_prl,Davidsen06_grl,Corral06_prl} were performed to characterize the spatio-temporal organization of earthquake events. In particular, Bak \etal (2002) \cite{Bak02_prl} showed that the distribution of waiting times $P(\Delta t)$ between earthquakes of energy larger than $E$ within a zone of size $L$ obeys to the following generalized expression of Omori's law:

\begin{equation}
P_{E,L}(\Delta t) = \Delta t^{-\alpha}f(\Delta t/\Delta t_0) \quad \mathrm{with} \quad \Delta t_0 \propto 1/(E^{-\beta} L^{d_f}),
\label{Sec:3:equ2}
\end{equation}

\noindent where $\beta$ is the exponent defined in Eq. \ref{Sec:3:equ1}, $\alpha \simeq 1$ is the Omori exponent, $d_f \simeq 1.2$ is a fractal dimension characterizing the spatial distribution of epicenters and $f(x)$ is a function constant for $x\ll 1$, and decaying rapidly with $x$ for $x \gg 1$. More recently, Davidsen and Paczuski (2005) showed \cite{Davidsen05_prl} that the spatial distance $\Delta r$ between the epicenters of two successive earthquakes of energy larger than $E$ within a zone of size $L$ exhibits scale-free statistics: 

\begin{equation}
P_{E,L}(\Delta r) = \left(\Delta r/L\right)^{-\delta} f(\Delta r/L), 
\label{Sec:3:equ3}
\end{equation}
\noindent where $\delta\simeq 0.6$ and $f(x)$ is a function constant for $x\ll 1$, and decaying rapidly with $x$ for $x \gg 1$.  

Since these scale-free distributions are observed universally on Earth, independently of the considered area of investigation and the period over which the analysis was performed, it was suggested \cite{Scholz68a_bssa} that similar statistical features should be reproducible in laboratory experiments. In this context, time series of {\em Acoustic Emission} (AE) were recorded in various rocks loaded under uniaxial compression up to (shear) fracture \cite{Scholz68a_bssa,Scholz68b_bssa,Hirata87_jgr,Davidsen07_prl}. As a result, the waiting time distribution $P(\Delta t)$ was found \cite{Davidsen07_prl} to obey Omori's law given by equation \ref{Sec:3:equ2} with an exponent $\alpha \simeq 1$ very close to that observed in earthquakes. 

\subsection{Acoustic emissions in tensile failure}\label{Sec:3.2}

Signature of crackling dynamics is also evidenced in the AE that goes with the mode I failure of many brittle materials: The distributions of energy and silent time between two successive events have been computed in many fracture experiments \cite{Garcimartin97_prl,Guarino98_epjb,Salminen02_prl,Davidsen07_prl,Koivisto07_prl,Deschanel06_ijf,Deschanel08_jsm} and found to follow power-laws similar to what observed in earthquakes (See equations \ref{Sec:3:equ1} and \ref{Sec:3:equ2}). It should be emphasized however that the relation between AE energy and released elastic energy remains largely unknown (see e.g. \cite{Minozzi03_epjb} for recent work in this context). The exponents $\beta_{AE}$ and $\alpha$ associated with AE energy and silent time were reported to depend on the considered material: They were found to be e.g. $\{\beta_{AE} \simeq 1.5, \alpha \simeq 1.9\}$ in wood \cite{Garcimartin97_prl,Guarino98_epjb}, $\{\beta_{AE} \simeq 2, \alpha \simeq 2.7\}$ in fiberglass \cite{Garcimartin97_prl,Guarino98_epjb}, $\{\beta_{AE} \simeq 1.3-1.8, \alpha \simeq 1-1.5 \}$ in polymeric foams \cite{Deschanel06_ijf,Deschanel08_jsm}\footnote{In polymeric foams, both $\beta_{AE}$ and $\alpha$ were also found to depend on the loading conditions (load-imposed or displacement-imposed) and on the temperature \cite{Deschanel06_ijf,Deschanel08_jsm}}, $\{\beta_{AE} \simeq 1,\alpha \simeq 1.25\}$ in sheets of paper \cite{Salminen02_prl}. As a synthesis, the values for $\beta_{AE}$ and $\alpha$ measured in mode I fracture experiments range typically between 1 and 2. Santucci \etal (2004) \cite{Santucci04_prl,Santucci07_epj} have also observed directly the slow sub-critical intermittent crack growth in 2D sheets of paper, up to their final breakdown. In these experiments, the crack tip is found to progress through discrete jump the size of which is power-law distributed with an exponent close to $1.5$ up to a stress-dependent cut-off that diverges as a power-law at $K_c$. The waiting time $\Delta t$ between two jumps is found to exhibit power-law distribution with Omori's exponent $\alpha$ close to $0.7$ for $\Delta t$ smaller than the mean value, and $\alpha=2$ for $\Delta t$ larger than the mean value. 

It is worth to notice that most of AE fracture experiments reported in the literature are non-stationary: Usually, one starts with an intact specimen and loads it up to its catastrophic failure. In these tests, the recorded AE reflects more the micro-fracturing processes {\em preceding} the initiation of the macroscopic crack than the growth of this latter. More recently, Salminen, Koivisto and co-workers (2006) \cite{Salminen06_epl,Koivisto07_prl} investigated the AE statistics {\em in a steady regime} of crack propagation in experiments of paper peeling. The distribution of AE energy follows a power-law with an exponent $\beta_{AE} \simeq 1.8\pm 0.2$ significantly higher than that observed in standard tensile tests starting from an initially intact paper sheet \cite{Salminen02_prl}. Nevertheless, the silent time between successive events was found to display statistics and correlations significantly more complex than that of Omori'law.        

\subsection{Intermittent crack growth along planar heterogeneous interfaces}\label{Sec:3.3}

\begin{figure}
\begin{center}
\includegraphics[width=0.8\textwidth]{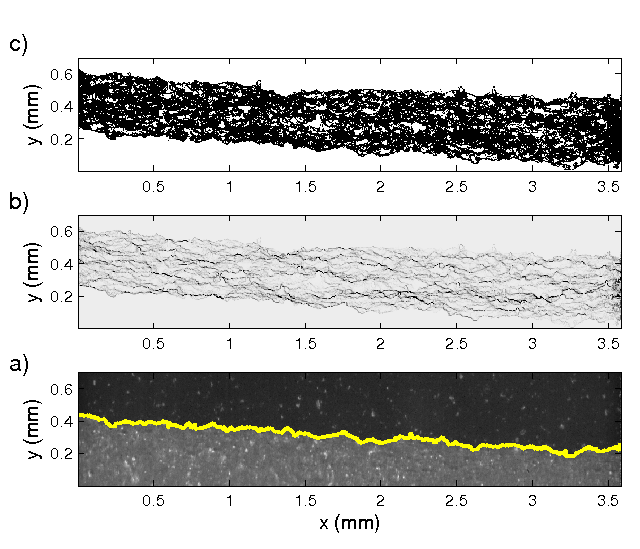}
\caption{(a) Typical image recorded by M{\aa}l{\o}y \etal of the crack front propagating along the weak heterogeneous interface between two sealed Plexiglas plates. The solid line represents the interface separating the uncracked (in black) and cracked (in gray) parts extracted after image analysis. (b) Gray scale map of the waiting time matrix. The darker parts show the longer waiting times.
(c) Spatial distribution of quakes (in white) corresponding to clusters with waiting time 10 times smaller than the mean one. (Courtesy of K.-J. M{\aa}l{\o}y, from \cite{Maloy06_prl}).}
\label{Sec:3:fig2}
\end{center}
\end{figure}

AE methods are very powerful to localize micro-fracturing events in time. On the other hand, to relate quantitatively the AE energy to the local crack dynamics or the mechanical energy released through these events remains a rather difficult task (see e.g. \cite{Minozzi03_epjb} for recent work in this context). This has motivated M{\aa}loy, Santucci and co-workers to investigate experimentally the dynamics of crack propagation in a simpler configuration, namely the one of a planar crack propagating along the weak heterogeneous interface between two sealed transparent Plexiglas plates \cite{Maloy01_prl,Maloy06_prl}. Using a fast camera, they observe directly crack propagation (Figure \ref{Sec:3:fig2}a, from \cite{Maloy06_prl}). 

To characterize the local dynamics, they computed the time spent by the crack front at each location $(z,x)$ of the recorded region. A typical gray-scale image of this so-called {\em waiting-time map}, $w(z,x)$, is shown in Figure \ref{Sec:3:fig2}b (from \cite{Maloy06_prl}). The intermittency reflects in the numerous and various regions of gray levels. The local speed $v(z,x)$ of the crack front as it passes through a particular location $(z,x)$ is then proportional to $1/w(z,x)$. Quakes are then defined \cite{Maloy06_prl} as connected zones where $v(z,x) \geq c \langle v \rangle$ where $\langle \rangle$ denotes averaging over all locations and $c$ is a constant ranging from 2 to 20 (Figure \ref{Sec:3:fig2}c, from \cite{Maloy06_prl}). Recent analyses performed by Grob \etal (2009) \cite{Grob09_pag} reveal many similarities between the statistics of these experimental quakes and that of real earthquakes: First, the distribution of quake area, $S_w$, follows a power-law, characterized by an exponent $\tau_w\simeq 1.7$ \cite{Maloy06_prl}, which was suggested \cite{Grob09_pag} to be analogous to the energy distribution characterized by an exponent $\beta \simeq 1.7$ observed in real earthquakes (Equation \ref{Sec:3:equ1}); Second, the distribution of time occurrence and spatial distance between the epicenters of two successive events were found to be the same in both experimental quakes and real earthquakes (Equations \ref{Sec:3:equ2} and \ref{Sec:3:equ3}, respectively). All these distributions were found to be independent of the clipping parameter $c$ used to define the quakes in the waiting-time maps \cite{Maloy06_prl,Grob09_pag}.

\begin{figure}
\begin{center}
\includegraphics[width=0.6\textwidth]{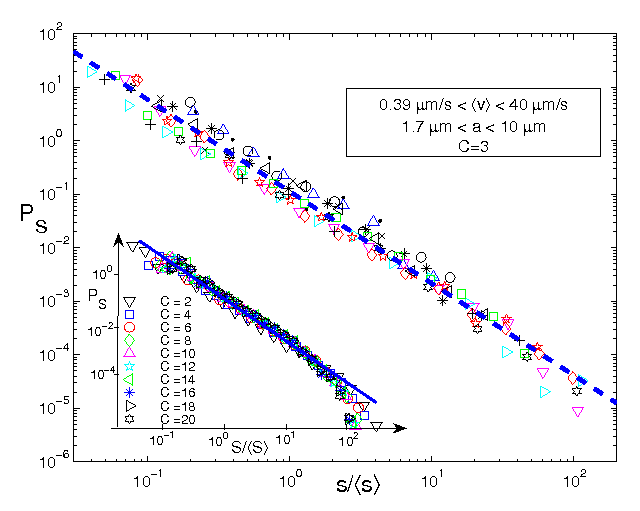}
\caption{Distribution $P(S_w)$ of the area of the quakes detected on the waiting time maps computed in the model experiment of M{\aa}l{\o}y \etal (see Figure \ref{Sec:3:fig2}). The notation $S \equiv S_w$ was the one used in ref. \cite{Maloy06_prl}.  The various symbols correspond to various values of mean front velocity $\langle v \rangle $ and pixel resolution $a$. Inset: Distribution of quake area for a wide range of clipping parameter $c$. The axes are logarithmic. In both plots, the straight line corresponds to a power-law distribution with $\tau_w \simeq 1.7$. (Courtesy of K.-J. M{\aa}l{\o}y, from \cite{Maloy06_prl}).}
\label{Sec:3:fig3}
\end{center}
\end{figure}

Let us draw some main conclusions from the experimental and fields' observations reviewed here:
\begin{itemize}
\item[(i)] Failure of brittle heterogeneous materials displays crackling dynamics, with sudden random events of energy release and/or jumps in the crack growth. The event statistics is characterized by various power-law distributions in e.g. energy, size or silent time between successive events. This crackling dynamics is observed in the micro-fracturing processes occurring prior to the initiation of a macroscopic crack {\em as well as} during the slow propagation of this latter.   
\item[(ii)] The exponents associated with the distribution of energy and silent times between the micro-fracturing events {\em preceding the initiation of a macroscopic crack} are generally observed to {\em depend on} the material, loading conditions, environment parameters...
\item[(iii)] The intermittency observed {\em during} slow crack propagation in an heterogeneous brittle medium loaded under tension shows strong similarities with the earthquake dynamics of faults. The exponents associated with the different power-law distributions seem to take universal values. 
\end{itemize}

Finally, it is worth to mention that crackling dynamics has also been observed, at the nanoscale, in cleavage experiments performed on pure crystals like mica \cite{Marchenko06_apl} and sapphire \cite{Astrom06_pla,Astrom07_condmat}. In the second case, both the distributions of released energy and silent time between two successive nanofracture events have been computed. They are found to obey power-law similar to that observed in AE during the fracture of brittle heterogeneous materials or in earthquakes.

\section{morphology of cracks}\label{Sec:4}

\begin{figure}
\begin{center}
\includegraphics[width=0.49\textwidth]{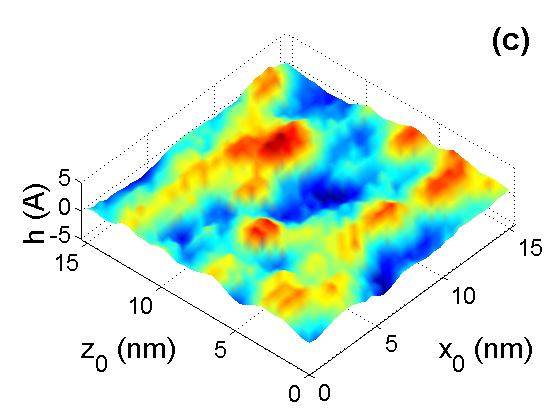}
\includegraphics[width=0.49\textwidth]{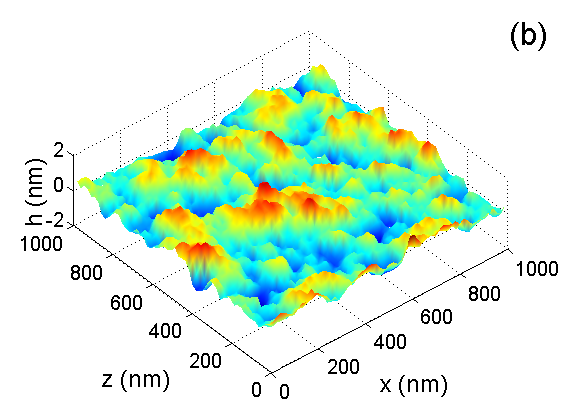}
\includegraphics[width=0.49\textwidth]{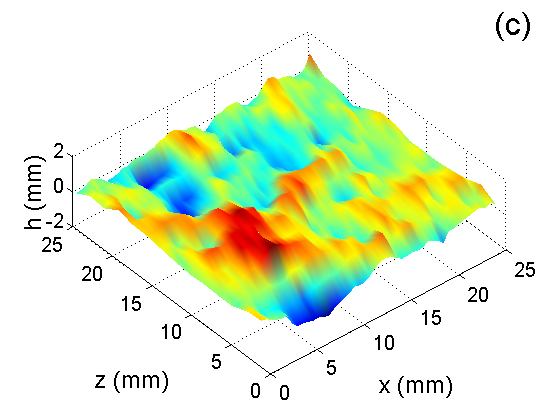}
\includegraphics[width=0.49\textwidth]{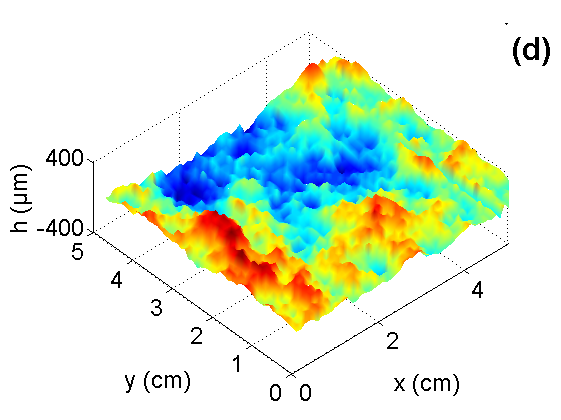}
\caption{Typical fracture surfaces in (a) quasi-crystals (STM, Courtesy of P. Ebert, from \cite{Ponson06_prb}), (b) glass (AFM, from \cite{Ponson06_prl}), (c) mortar (Contact profilometry, courtesy of S. Morel, from \cite{Ponson06_ijf}) and (d) glassy ceramics made of sintered glass beads (Interferometric profilometry). In all these materials, the roughness extends over length-scales much larger than that of the microstructure.}
\label{Sec:4:fig1}
\end{center}
\end{figure}

We turn now to crack roughness. The morphology of fracture surfaces has been widely investigated over the last century - fractography is now routinely used to determine the causes of failure in engineering structures (see e.g. \cite{Hull99_book} for a recent review). In section \ref{Sec:2.3}, we saw that LEFM states that crack propagation in brittle materials remains in pure tension as long as $v$ remains small. This would lead to smooth fracture surfaces. Experimentally, the failure of disordered media leads to rough surfaces at length scales much larger than that of the microstructure (Figure \ref{Sec:4:fig1}).

\subsection{Self-affine scaling features of crack profiles}\label{Sec:4.1}

\begin{figure}
\begin{center}
\includegraphics[width=0.95\textwidth]{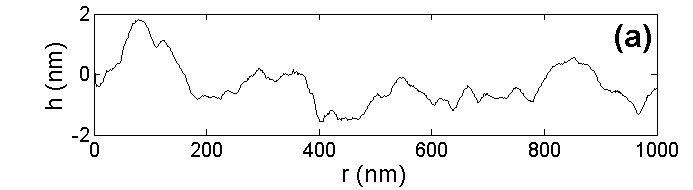}
\includegraphics[width=0.49\textwidth]{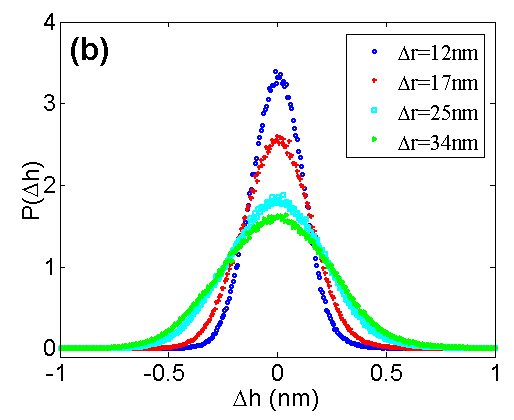}
\includegraphics[width=0.49\textwidth]{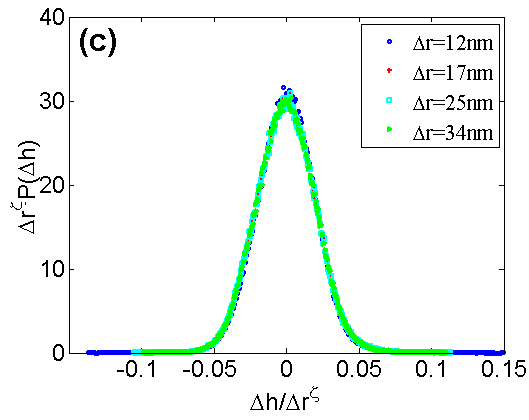}
\caption{(a) Typical fracture profile as observed via AFM in silica glass broken under stress corrosion. This profile was taken perpendicular to the direction of crack growth. (b) Distribution of the increments $\Delta h$ between two points of such profiles for various values of $\Delta r$. (c) Data collapse of this distribution using Eq. \ref{Sec:4:equ1} with $\zeta \simeq 0.8$ which shows that the profile is self-affine with a roughness exponent $\zeta=0.8$.}
\label{Sec:4:fig2}
\end{center}
\end{figure}

Starting from the pioneer work of Mandelbrot \etal (1984) \cite{Mandelbrot84_nature}, many experiments have revealed that fracture surfaces exhibit self-affine morphological scaling features. In other words, crack profiles such as the one shown in Figure \ref{Sec:4:fig2}a are statistically invariant through the affine transformation $(r,h) \rightarrow (\lambda r,\lambda^\zeta h)$ where $r$ and $h(r)$ refer to the in-plane and out-of-plane coordinates respectively. The exponent $\zeta$ is called the {\em Hurst or roughness exponent}. Experimentally, this affine invariance can be checked by computing the distribution $P_{\Delta r} (\Delta h)$ of height increments between two points separated by a distance $\Delta r$ (Figure \ref{Sec:4:fig2}b): For self-affine crack profiles, $P_{\Delta r} (\Delta h)$ takes the following form (Figure \ref{Sec:4:fig2}c): 

\begin{equation}
P_{\Delta r}(\Delta h) = \Delta r^{-\zeta} f(\Delta h / \Delta r^ \zeta) 
\label{Sec:4:equ1}
\end{equation}

The shape of the generic distribution $f$ is often found to be very close to a Gaussian (see e.g. \cite{Santucci07_pre,Ponson07_pre} and Figure \ref{Sec:4:fig2}c). Then, $\zeta$ allows one to characterize entirely the statistics of crack roughness. In this respect, its determination was an important issue and lead to many studies over the last 25 years. Besides the full computation of the distribution $P_{\Delta r}(\Delta h)$ that presents the drawback to necessitate a large amount of data, several other methods were developed to measure $\zeta$. They will not be reviewed here. An interested reader may refer to the work of Schmittbuhl (1995) \cite{Schmittbuhl95_pre} for more information on these methods and their limitations. In the following, we will use one of the most common ones, which consists in computing the height-height correlation function:

\begin{equation}
\Delta h(\Delta r) = \langle (h(r+\Delta r)-h(r))^2 \rangle _r^{1/2}, 
\label{Sec:4:equ2}
\end{equation}

\noindent where $\langle \rangle_r$ denotes average over all positions $r$ along a given profile. For a self-affine profile, one gets:

\begin{equation}
\Delta h(\Delta r) \propto \Delta r^{\zeta} 
\label{Sec:4:equ3}
\end{equation}

\noindent It is important to emphasize that this method, - as well as many others -, allows to measure the roughness exponent {\em provided that the profile is actually self-affine}, but do not ensure this self-affinity by itself. 

Possible universality in the morphological scaling features of cracks was first mentioned by Termonia and Meakin (1986) \cite{Termonia86_nature}: They made use of a minimal 2D molecular model of material failure to address the question and observed that the fractal dimension of the final crack was independent of the elastic constants over a wide range of values. Bouchaud \etal (1990) \cite{Bouchaud90_epl} were the first to suggest a universal value for the roughness exponent $\zeta$ on the basis of experimental observations: In their seminal series of fractography experiments, the authors observed that $\zeta \simeq 0.8$ in four specimens of aluminium alloy on which different heat treatments had conferred different microstructures and toughness. This universality seemed to be corroborated by Maloy et al (1992) \cite{Maloy92_prl} who observed similar $\zeta$ values in six different brittle materials. Thereafter, many experimental works were focused on the measurement of $\zeta$ - the review written by Bouchaud (1997) \cite{Bouchaud97_jpcm} synthesizes most of them up to 1997. Reported $\zeta$ values are between $0.5$ and $0.9$ in metallic alloys \cite{Mandelbrot84_nature,Bouchaud90_epl,Millman94_pms,Morel04_prl}, around $0.9$ in graphite \cite{Maloy92_prl}, around $0.75$ in porcelain \cite{Maloy92_prl}, around $0.85-0.9$ in Bakelite \cite{Maloy92_prl}, between $0.5$ and $0.8$ in oxide glasses \cite{Daguier97_prl,Hinojosa08_ijf}, around $0.8-0.9$ in wood \cite{Morel98_pre} and in ice \cite{Weiss02_efm}, around $0.45-0.5$ in Sandstone \cite{Boffa98_epjap,Boffa00_pa,Ponson07_pre},around $0.45-0.5$ in Basalt \cite{Boffa00_pa}, around $0.7-0.8$ in mortar \cite{Mourot05_pre}, and around $0.4-0.5$ in glassy ceramics made of sintered glass beads with various porosities \cite{Ponson06_prl2,Ponson07_AnPhys}. As a result, these fractography experiments performed over a period spreading from 1984 to 2006 leaded to mitigated opinions on the relevance or not of this universality \cite{Millman94_pms,Bouchbinder05_prl}. As we will see in the two next sections, some recent observations allow to reconciliate this apparent spreading in the measurements of $\zeta$ with the scenario of universal morphological scaling features.

All the measurements reported in the preceding paragraph concern the roughness of fracture surfaces resulting from the failure of 3D specimens. Crack morphology was also explored in 2D geometries. Crack lines in paper broken in tension were also found to be self-affine with a roughness exponent $\zeta \simeq 0.7$ \cite{Kertesz93_fractals,Salminen03_epjb,Santucci07_pre}. Note that a small but clear difference was found in $\zeta$ between the slow (subcritical) and the fast growth regime \cite{Mallick07_prl}. Crack front in the interfacial fracture experiments presented in section \ref{Sec:3.3} were also reported to exhibit self-affine scaling features. The associated roughness exponent was found to be $\zeta_{H} \simeq 0.6$ \cite{Schmittbuhl97_prl,Delaplace99_pre}. 

Let us end this section by mentioning that this picture of simple self-affine cracks was recently questioned. First, {\em multi-scaling}, characterized by non-constant scaling exponents $\zeta_q$ between the higher order height-height correlation function $\Delta h_q(\Delta r) = \langle (h(r+\Delta r)-h(r))^q \rangle _r^{1/q}$ and $\Delta r$, was invoked in both 2D geometries \cite{Bouchbinder06_prl}, and 3D geometries \cite{Schmittbuhl95_jgr}. This multi-scaling seems however to disappear at large scales \cite{Salminen03_epjb,Alava06_jsm,Santucci07_pre}. Second, fracture surfaces exhibit {\em anomalous scaling} \cite{Lopez98_pre,Morel98_pre}: The introduction of an additional {\em global} roughness exponent, $\zeta_{glob}$, is necessary to describe the scaling between the global crack width and the specimen size. Experiments reveal that $\zeta_{glob}$ depends on the material and on the fracture geometry \cite{Lopez98_pre,Morel98_pre}.

\subsection{Family-Viseck scaling of fracture surfaces}\label{Sec:4.2}

\begin{figure}
\begin{center}
\includegraphics[width=0.49\textwidth]{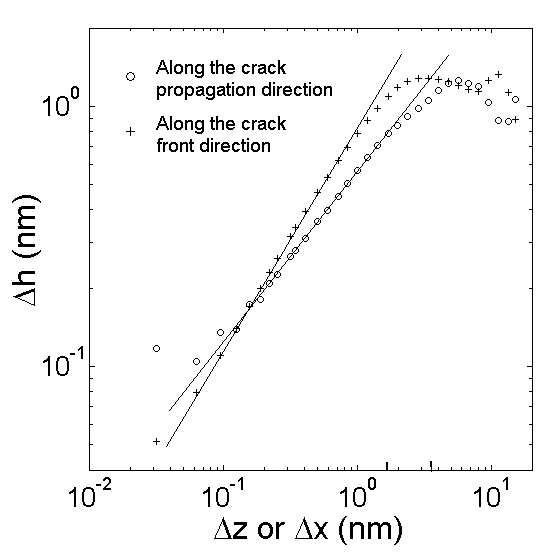}
\includegraphics[width=0.49\textwidth]{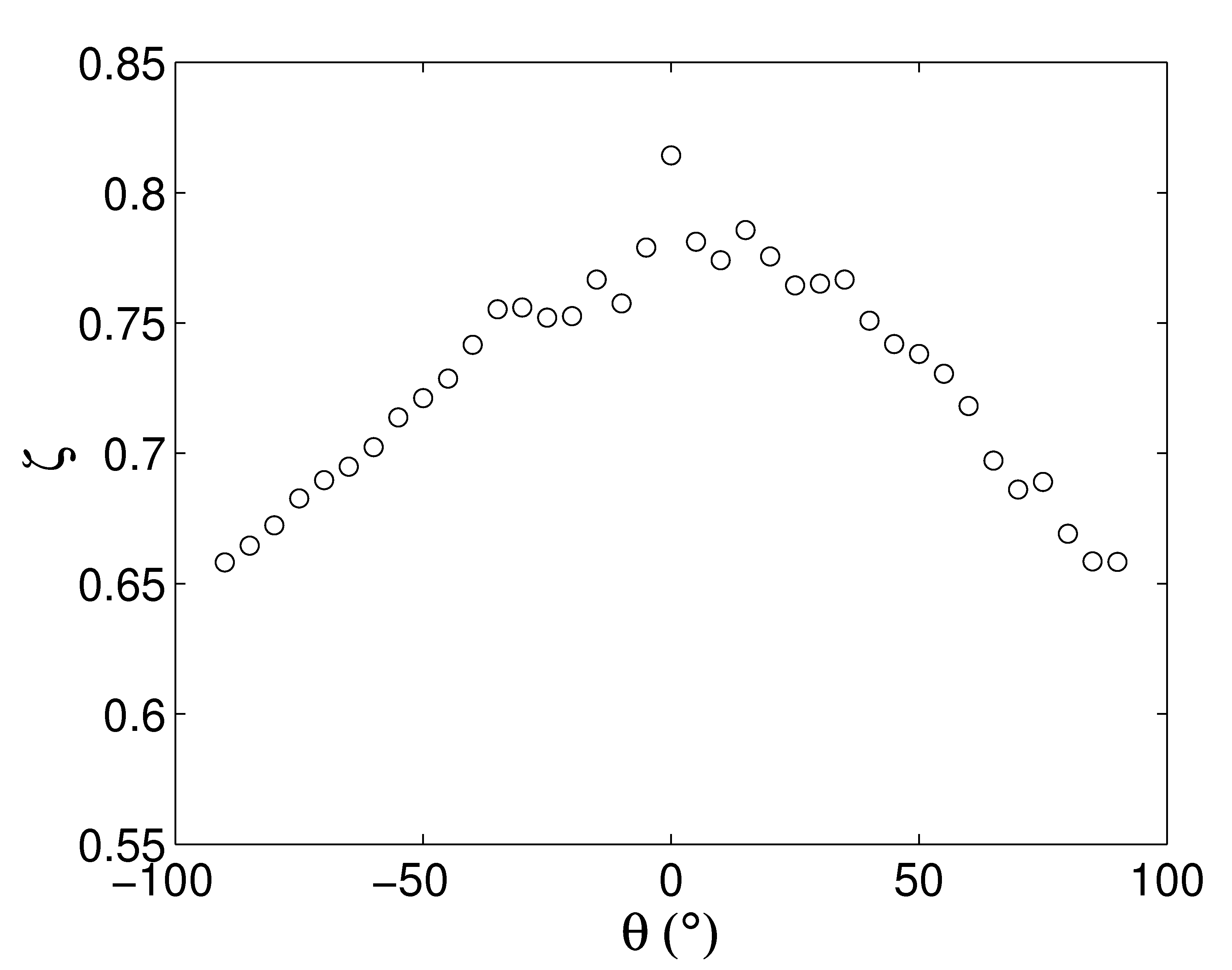}
\caption{(a) 1D height-height correlation function measured parallel and perpendicular to the crack propagation direction. The axes are logarithmic. (b) Variation of roughness exponent $\zeta$ as a function of the angle $\theta$ between the considered profile and the direction of crack propagation. Data were measured in quasi-crystals (From \cite{Ponson06_prb}).}
\label{Sec:4:fig3}
\end{center}
\end{figure}

To uncover the primary cause of the apparent spreading in the measured roughness exponents, scaling anisotropy were looked for \cite{Ponson06_prl,Ponson06_prb,Ponson06_ijf} in the fracture surfaces of various materials (quasi-crystals, glasses, metallic alloys, wood and mortar) broken under various conditions (cleavage, stress corrosion, tension) and various velocities, ranging from $~10^{-12}\un{m/s}$ to $~10^{2}\un{m/s}$. At this point, it is important to set the notations (Figure \ref{Sec:6:fig1}): In all the following, the axis $\vec{e}_x$, $\vec{e}_y$ and $\vec{e}_z$ will refer to the direction of crack propagation, the direction of tensile loading and the mean direction of the crack front, respectively. Special attention was paid to ensure that fracture surfaces were observed in a steady state regime, far enough from crack initiation so that roughness becomes statistically invariant of the position $x$ along the direction of crack growth. By computing the height-height correlation function $\Delta h_{\theta}(\Delta r)$ for profiles making various angles $\theta$ with $\vec{e}_x$, the existence of anisotropy in the scaling features (Figure \ref{Sec:4:fig3}) was evidenced: $\zeta(\theta)$ varies from $\zeta(\theta=0)=\zeta_{||} \simeq 0.6$ (profiles taken {\em parallel} to the direction of crack growth) to $\zeta(\theta=90^\circ)=\zeta_{\perp} \simeq 0.8$ (profiles taken {\em perpendicular} to the direction of crack growth, parallel to the mean direction of crack front).

As a consequence, a full characterization of the scaling statistical properties of fracture surfaces calls for the computation of the 2D height-height correlation function:

\begin{equation}
\Delta h(\Delta z,\Delta x)=\langle (h(z+\Delta z,x+\Delta x)-h(z,x))^2\rangle _{z,x}
\label{Sec:4:equ4}
\end{equation}

\noindent This was shown \cite{Ponson06_prl,Ponson06_prb,Ponson06_ijf} to take a peculiar form, referred to as Family-Viseck scaling \cite{Family91_book}:

\begin{equation}
\begin{array} {l}
   \Delta h(\Delta z,\Delta x)\propto \Delta x^{\zeta_{||}}f(\Delta z/\Delta x^{\zeta_{||}/\zeta_{\perp}}) \\
\\
$where$\quad	f(u) \propto \left\{
\begin{array}{l l}
1 & $if u$ \ll c  \\
u^{\zeta_{\perp}} & $if u$ \gg c
\end{array}
\right.
\end{array}
\label{Sec:4:equ5}
\end{equation}

Such Family-Vicsek scalings are classically observed in interface growth processes. In this respect, the exponent $\zeta_{||}$and the ratio $\zeta_{\perp}/\zeta_{||}$ can be mapped to a {\em growth} exponent and a {\em dynamic} exponent, respectively. It is worth to mention that such an anisotropic Family-Viseck scaling of fracture surfaces was first suggested by the phenomenological Langevin description of crack propagation proposed by Bouchaud \etal (1993) \cite{Bouchaud93_prl} detailed thereafter in section \ref{Sec:6.3}.

\begin{figure}
\begin{center}
\includegraphics[width=0.49\textwidth]{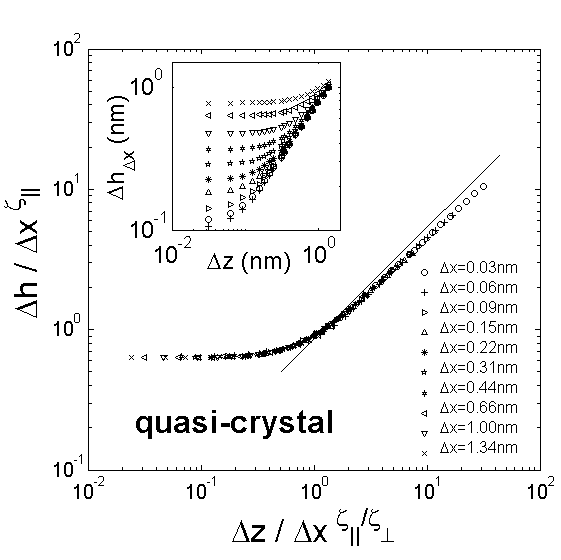}
\includegraphics[width=0.49\textwidth]{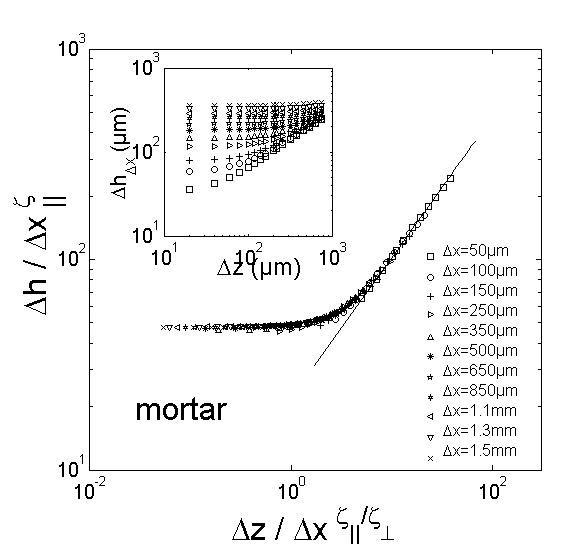}
\caption{Family-Viseck scaling of the 2D height-height correlation function in quasi-crystals (From \cite{Ponson06_prb}) and in mortar (From \cite{Ponson06_ijf}) using equation \ref{Sec:4:equ5}.}
\label{Sec:4:fig4}
\end{center}
\end{figure}

The two exponents $\zeta_{\perp}=0.75\pm 0.05$ and $\zeta_{||}=0.6\pm0.05$ were found to be universal, independent of the considered material, the failure mode and the crack growth velocity over the whole range from ultra-slow stress corrosion fracture (picometer per second) to rapid failure (few hundreds meter per second). On the other hand, the range of length-scales over which these Family-Viseck scaling features are observed is limited and material dependent. This will be discussed in section \ref{Sec:4.3}.

By rescaling the distances $\Delta x$ and $\Delta z$ by the topothesies $\ell_x$ and $\ell_z$ defined as the scales at which the out-of-plane increment becomes equal to the in-plane one, i.e.  $\Delta h(\Delta z=0,\Delta x=\ell_x)=\ell_x$ and $\Delta h(\Delta z=\ell_z,\Delta x=0)=\ell_z$, one can rewrite the Family-Viseck scaling \cite{Ponson06_prb}:

\begin{equation}
\begin{array} {l}
\Delta h(\Delta z,\Delta x) =\ell_x (\Delta x/\ell_x )^{\zeta_{||}} g \left( u= \frac{\ell_z}{\ell_x}\frac{\Delta z/\ell_z}{(\Delta x/\ell_x)^{\zeta_{||}/\zeta_{\perp}}} \right)\\
\\
\mathrm{where} \quad g(u) =
\left\{
\begin{array}{l l}
1 & $if u$ \ll 1  \\
u^{\zeta_{\perp}} & $if u$ \gg 1
\end{array}
\right.
\end{array}
\label{Sec:4:equ6}
\end{equation}
 
\begin{figure}
\begin{center}
\includegraphics[width=0.6\textwidth]{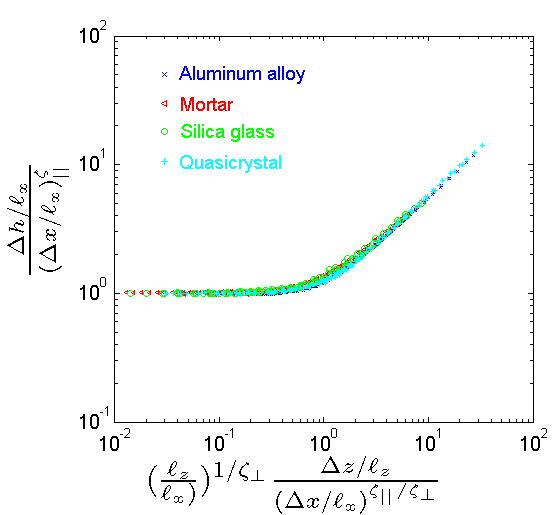}
\caption{Collapse of the dimensionless 2D height-height correlation function using Equation \ref{Sec:4:equ6} in various materials, (From \cite{Ponson06_prb}).}
\label{Sec:4:fig5}
\end{center}
\end{figure}

\noindent The form of $f$ is then found to be universal \cite{Ponson06_prb}, independent of the considered material, of the failure mode and of the crack growth velocity (Figure \ref{Sec:4:fig5}). 

It is important to insist that this universal Family-Viseck scaling is observed in fracture surfaces far enough from crack initiation, in a regime where roughness becomes statistically invariant with respect to translations along the direction of crack growth $\vec{e}_x$. 
The transient roughening regime starting from an initial straight notch, was extensively studied by Lopez, Morel \etal \cite{Lopez98_pre,Morel98_pre,Morel08_pre}. It was found to display a scaling significantly more complex than that expected in Family-Viseck scenario. It involves the global roughness exponent $\zeta_{glob}$ defined at the end of section \ref{Sec:4.1} as well as a new independent dynamic exponent. These two exponents are found to depend on both the material and specimen geometry.

\subsection{On the relevant length-scales}\label{Sec:4.3}

Several fractography observations reported in the literature are found not to be compatible with the Family-Viseck scaling (equation \ref{Sec:4:equ5}) and its associated exponents $\{\zeta_{\perp} \simeq 0.75,\zeta_{||}\simeq 0.6 \}$. In particular, experiments in sandstone \cite{Boffa98_epjap,Boffa00_pa,Ponson07_pre}, in artificial rocks \cite{Bouchbinder06_prl} and in granular packings made of sintered glass beads \cite{Ponson06_prl2,Bonamy06_prl,Ponson07_AnPhys} have shown self-affine scaling properties characterized by roughness exponents significantly smaller, around $0.4-0.5$. In the case of granular packings made of sintered glass beads, the 2D height-height correlation function was found to exhibit Family-Viseck scaling as for the materials investigated in the previous section, but with $\{\zeta_{\perp} \simeq 0.4,\zeta_{||}\simeq 0.5\}$ \cite{Bonamy06_prl}. This suggests the existence of a second universality class for post-mortem fracture surfaces.  

\begin{figure}
\begin{center}
\includegraphics[width=0.6\textwidth]{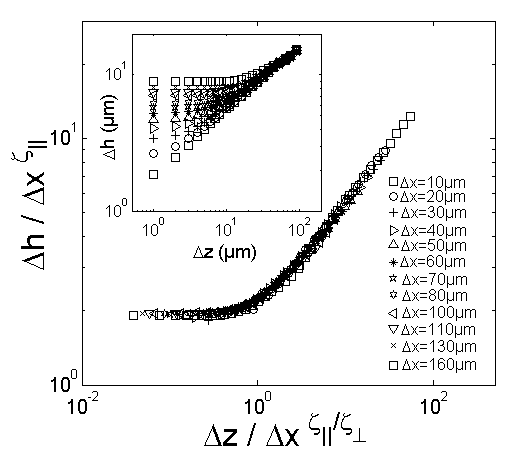}
\caption{Variation of the 2D height-height correlation function $\Delta h$ with $\Delta z$ for various values of $\Delta x$ for granular packings made of sintered glass beads. The inset shows the direct curves while the data collapse in the main graph was obtained from equation \ref{Sec:4:equ5} using $\zeta_{\perp}=0.4$ and $\zeta_{||} =0.5$ (From \cite{Bonamy06_prl}).}
\label{Sec:4:fig6}
\end{center}
\end{figure}

To uncover the origin of these two distinct series of observed exponents, The range of length-scales over which these two regimes are observed was examined \cite{Bonamy06_prl}. In all experiments compatible with $\{\zeta_{\perp} \simeq 0.8,\zeta_{||}\simeq 0.6\}$, fracture surfaces were observed at {\em small} length-scales, below a given cutoff length $\xi$ of e.g. few nanometers in quasicrystals \cite{Ponson06_prb}, few tens of nanometers in glass \cite{Daguier97_prl,Bonamy06_prl,Dalmas08_prl}, few hundreds of micrometers in metallic alloys \cite{Daguier97_prl,Morel04_prl}, few millimeters in wood \cite{Morel98_pre,Ponson06_prl}, few tens of millimeters in mortar \cite{Mourot05_pre,Morel08_pre}... On the other hand, observations compatible with $\{\zeta_{\perp} \simeq 0.4,\zeta_{||}\simeq 0.5\}$ were observed at {\em large} scales, well above the microstructure scale, up to a cutoff length set by the specimen size \cite{Ponson07_pre}.

It was then proposed, in ref. \cite{Bonamy06_prl}, that the upper cutoff length $\xi$ that limits the $\{\zeta_{\perp} \simeq 0.8,\zeta_{||}\simeq 0.6\}$ Family-Viseck scaling regime is set by the size of FPZ. This conjecture was proven to be true in oxide glasses \cite{Bonamy06_prl}, quasicrystals \cite{Ponson06_prb} and mortar \cite{Morel08_pre}. Work in progress \cite{Guerra08_unp} seems to indicate that it is also true in metallic alloys: CT specimen were broken at various temperatures, ranging from $20\un{K}$ to $148\un{K}$ - this allows to tune the FPZ size, from $20~{\mu}\mathrm{m}$ to $1\un{mm}$. And in all these tests, the cutoff length $\xi$ of the self-affine $\zeta \simeq 0.8$ regime is found to be roughly proportional to the process zone size.

It is worth to mention that no anomalous scaling was reported in Sandstone fractography experiments \cite{Ponson07_pre}. This suggests that the anomalous scaling commonly associated with the $\zeta \simeq 0.8$ self-affine regime \cite{Lopez98_pre,Morel98_pre} is absent in the large scale $\zeta \simeq 0.4$ self-affine regime. It is then appealing to associate the observation of this anomalous scaling to damage spreading at scale smaller than the FPZ.

Let us note also that a {\em third} isotropic scaling regime arising at {\em small scale}, characterized by a roughness exponent $\zeta \simeq 0.5$ was observed in {\em ductile} materials like Ti$_3$Al-based super-$\alpha_2$ intermetallic alloy broken under fatigue within $1\un{nm}-10~\mu\mathrm{m}$ range \cite{Daguier97_prl,Bouchaud97_jpcm,Bouchaud03_srl}, and in Zr-based metallic glass within the range $100\un{nm}-1\un{mm}$ \cite{Bouchaud08_epl}. 
It should be emphasized that in the second case, the distribution of height increments does not display the collapse given by equation \ref{Sec:4:equ1} expected for truly self-affine profiles. Similar $\zeta \simeq 0.5$ isotropic regime was also claimed \cite{Daguier97_prl,Bouchaud97_jpcm} to be observed on the AFM fracture surfaces of soda-lime broken under stress corrosion, within $1\un{nm}-10\un{nm}$ range. However, this observation has been recently called into question since the observations remain confined within (in-plane) AFM resolutions. Moreover, it was not reproduced in nanoresolved fracture surfaces of various silicate glasses (silica, soda-lime, borosilicate and aluminosilicate) broken under stress corrosion with crack velocities as small as the picometer per second \cite{Ponson06_prl,Bonamy06_prl}. As a result, this third small-scale regime characterized by $\zeta \simeq 0.5$ is suspected to be inherent to ductile fracture. As such, it will not be discussed in the sequel.

Let us conclude this section by drawing the outlines of the fractography experiments reviewed here:
\begin{itemize}
\item[(i)] Failure of brittle materials leads to rough fracture surfaces, with self-affine morphological features. 
\item[(ii)] Far enough from crack initiation, the morphology of fracture surfaces exhibits anisotropic Family-Viseck scaling, characterized by {\em two} distinct roughness exponents $\zeta_{||}$ and $\zeta_{\perp}$, measured parallel and perpendicular to the direction of crack propagation, respectively. By analogy with interface growth problems, the exponent $\zeta_{||}$ and the ratio $\zeta_{\perp}/\zeta_{||}$ can be mapped to a growth and a dynamic exponent, respectively.
\item[(iii)] {\em Two} distinct sets of universal exponents are observed: $\{\zeta_{\perp} \simeq 0.8,\zeta_{||}=0.6\}$ at small scales and $\{\zeta_{\perp} \simeq 0.4,\zeta_{||}=0.5\}$ at large scale.
\item[(iv)] The cutoff $\xi$ of the small-scale regime is set by the FPZ size. 
\end{itemize}

\section{Numerical observations in lattice models}\label{Sec:5}

The two preceding sections allowed to illustrate that the failure of brittle heterogeneous materials leads to (i) intermittent crackling dynamics and (ii) rough fracture surfaces. In both cases, one observes scaling invariance in term of distribution and morphology. Some of these scaling features appear universal - the associated scaling exponents are the same in a wide range of materials and loading conditions. This is reminiscent of critical phenomena where the balance between local (thermal or quenched) disorder and simple interactions between microscopic objects may select such complex scale invariant organization of the system (see e.g. \cite{Pfeuty77_book} for an introduction to critical phenomena). 

\begin{figure}
\begin{center}
\includegraphics[width=0.95\textwidth]{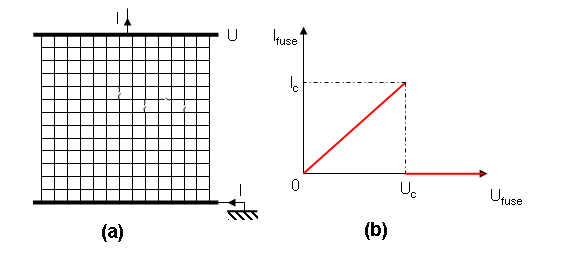}
\caption{(a) Random Fuse Networks with a square lattice geometry. The system is "loaded" either by imposing the voltage $U$ at the upper electrode (fracture test with imposed displacement conditions) or by imposing the total current $I$ crossing the system (fracture test with imposed force conditions). (b) Each fuse has a unity conductance and a randomly assigned breakdown threshold $\{U_c,I_c\}$.}
\label{Sec:5:fig1}
\end{center}
\end{figure}

In this context, it was proposed, initially by de Arcangelis \etal (1989) \cite{deArcangelis89_prb}, to exploit the analogy between scalar mode III elasticity and electricity and sketch an heterogeneous brittle material as a network of fuses of unit ohmic resistance and randomly distributed breakdowns (Figure \ref{Sec:5:fig1}). In these so-called Random Fuse Models (RFM), current and voltage are analogous to force and displacement, respectively. The goal is not so to reproduce exactly the failure of real brittle materials, but to see to which extent one can reproduce the scaling features presented in Sections \ref{Sec:3} and \ref{Sec:4} in such minimalist materials, keeping only the two main ingredients characterizing material failure, namely the microstructure randomness and the long-range coupling that accompanies the load redistribution after each element breakdown.      

\subsection{Intermittency}\label{Sec:5.1}

\begin{figure}
\begin{center}
\includegraphics[width=0.32\textwidth]{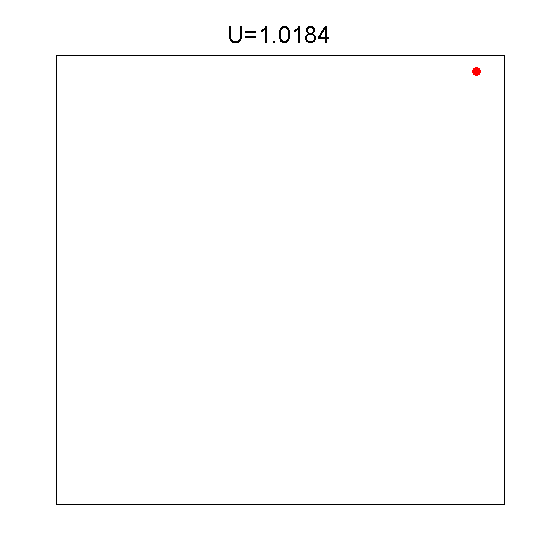}
\includegraphics[width=0.32\textwidth]{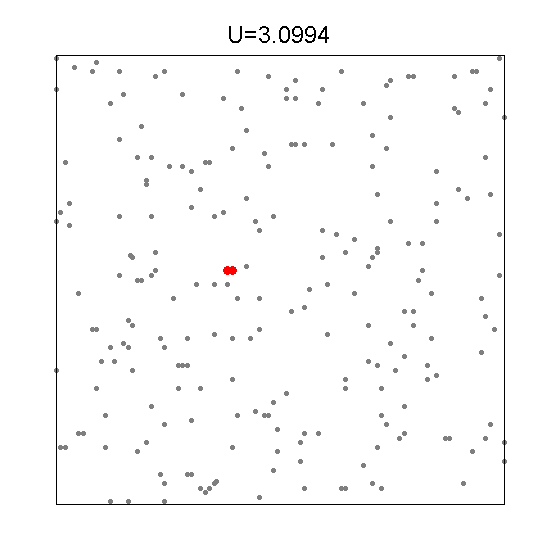}
\includegraphics[width=0.32\textwidth]{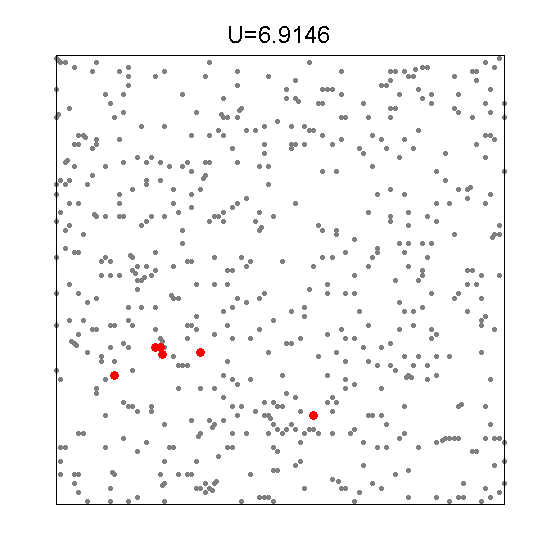}
\includegraphics[width=0.32\textwidth]{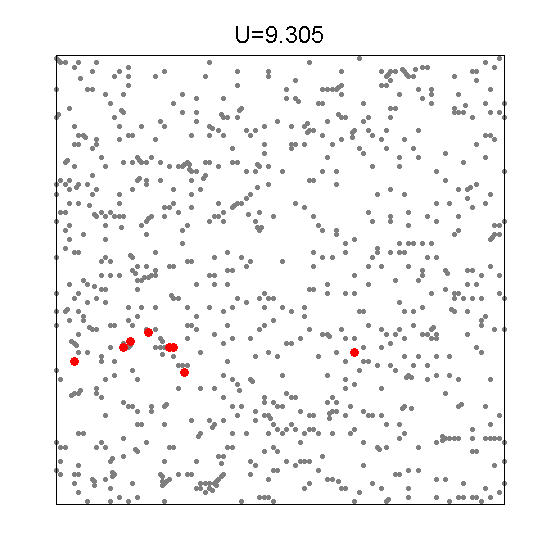}
\includegraphics[width=0.32\textwidth]{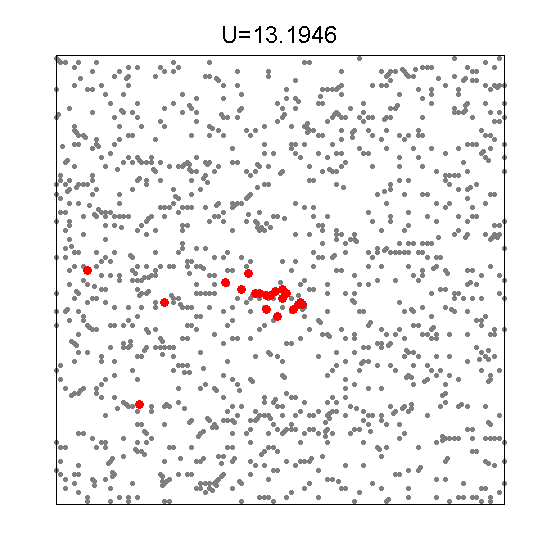}
\includegraphics[width=0.32\textwidth]{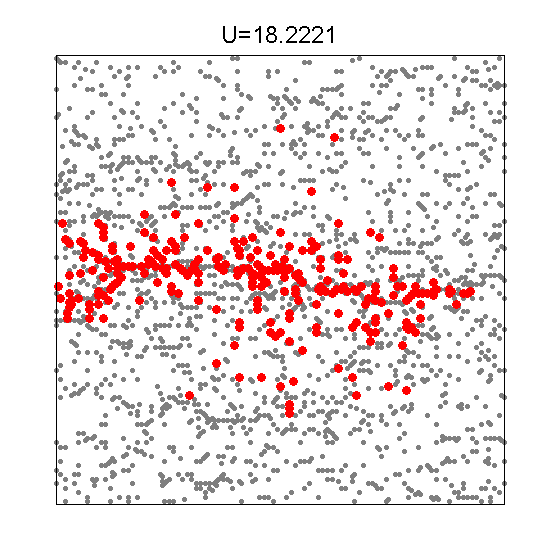}
\caption{Successive snapshots in a typical simulation of 2D $100\times 100$ RFM (square lattice) "loaded" by gradually increasing the voltage $U$. The red large points correspond to the cascade of fuses breakdown at the considered value of $U$. The gray small points corresponds to the fuses that have been burnt from $U=0$ to the considered value of $U$. In the last image of this sequence, the lattice is broken into two pieces. Note the roughness of the fracture line.}
\label{Sec:5:fig2}
\end{center}
\end{figure}

\begin{figure}
\begin{center}
\includegraphics[width=0.6\textwidth]{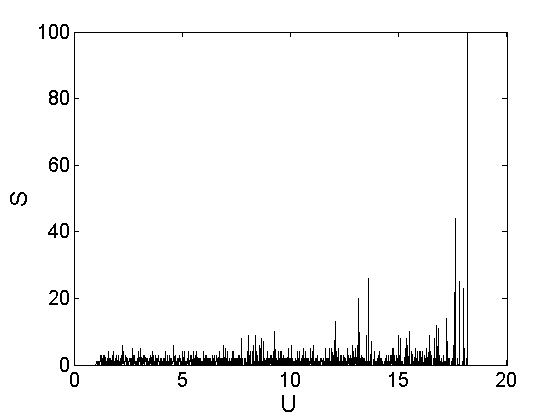}
\caption{Evolution of the avalanche size $S$, i.e. the number of fuse participating in a breakdown cascade as a function of the applied voltage $U$ in the RFM simulation presented in Figure \ref{Sec:5:fig2}.}
\label{Sec:5:fig3}
\end{center}
\end{figure}

A typical "fracture" experiment performed on 2D RFM is represented in Figure \ref{Sec:5:fig2}. The "loading" is increased by increasing the voltage $U$ of the nodes belonging to the upper electrode while keeping to zero the bottom ones. Voltages at each node and currents crossing each element are determined such that Kirchoff law and Ohm's law are satisfied everywhere. At some point, a fuse breaks. As a result, the current increases abruptly in the remaining elements. In turn, this may trigger a cascade, avalanche, of elements breakdowns that can either lead to a stable situation or to the failure of the whole structure. Typical evolution of the size of these avalanches as a function of the imposed voltage $U$ is plotted in figure \ref{Sec:5:fig3}. Avalanches are analogue to the AE events observed prior to the catastrophic failure of real brittle materials, as discussed in section \ref{Sec:3.2}.  

The distribution $P(S)$ of avalanche sizes, i.e. the number of fuses participating in a breakdown cascade has been investigated numerically, first by Hansen and Hemmer (1994) \cite{Hansen94_pl} and later by other teams \cite{Zapperi97_prl,Raisanen98_prb,Zapperi05_pre,Zapperi05_pa}. It was found to follow a power-law $P(S) \propto S^{-\tau}$ up to a cut-off $S_0$ that scales with the lattice size $L$ as $S_0 \propto L^D$ where $D$ refers to the avalanche fractal dimension. Initially, 2D simulations yielded a universal value $\tau \simeq 2.5$ \cite{Hansen94_pl,Zapperi97_prl}. However, recent large-scale simulations suggest that $\tau$ depends slightly on the lattice type in 2D systems: $\tau=2.75$ and $\tau=3.05$ in diamond and triangular lattice respectively \cite{Zapperi05_pre}. On the other hand, $D$ was found to be universal: $D\simeq 1.18$. In 3D simulations, both exponents were found to be universal \cite{Zapperi05_pa}: $\tau=2.5$ and $D=1.5$.

\subsection{Morphology of fracture surfaces}\label{Sec:5.2}

The morphology of final cracks has also been investigated numerically \cite{Hansen91_prl,Batrouni98_prl, Raisanen98_prl,Raisanen98_prb,Seppala00_pre,Zapperi05_pre,Nukala06_pre,Nukala07_pre}. As in experiments, it was found to exhibit self-affine scaling features. In 2D RFM such as the one shown in Figure \ref{Sec:5:fig2}, large scale simulations yield a universal roughness exponent $\zeta=0.71$ independent of the lattice topology, the presence or not of a pre-existing initial notch and of the disorder in breaking thresholds \cite{Zapperi05_pre,Nukala07_pre}. In 3D systems, $\zeta \simeq 0.42$ \cite{Raisanen98_prl,Raisanen98_prb,Nukala06_pre}.

It is interesting to note that the value of $\zeta$ measured in 2D RFM is pretty close to the $\zeta \simeq 0.7$ observed experimentally in quasi-two-dimensional materials \cite{Kertesz93_fractals,Salminen03_epjb,Santucci07_pre}. In 3D RFM, the value $\zeta \simeq 0.42$ are significantly smaller than the $\zeta \simeq 0.8$ observed experimentally in a wide range of materials (see sections \ref{Sec:4.1} and \ref{Sec:4.2}). On the other hand, it is pretty close to the value $\zeta \simeq 0.4 - 0.5$ observed at scale larger than the FPZ size in sandstone \cite{Boffa98_epjap,Ponson07_pre}, Glassy ceramics made of sintered glass beads \cite{Ponson06_prl2,Ponson07_AnPhys} and mortar \cite{Morel08_pre}. It should be emphasized that the morphological scaling analyses reported in \cite{Raisanen98_prl,Raisanen98_prb,Nukala06_pre} were performed by averaging together the results of different simulations starting from configurations without initial notch, and therefore no prescribed direction of crack propagation. As a consequence, the scaling anisotropy reported in experiments and described in section \ref{Sec:4.2} cannot be captured.

Let us finally mention that large scale numerical simulations revealed the existence of anomalous scaling in both 2D and 3D RFM \cite{Zapperi05_pre,Nukala07_pre}. The origin of this anomalous scaling remains unclear. It seems to be intrinsic to the scalar dimensionality of RFM since it disappears in Random Beam Models \cite{Nukala08_pre}. 

\section{Stochastic theory of crack growth}\label{Sec:6}

Experimental observations gathered in sections \ref{Sec:3} and \ref{Sec:4} revealed that slow crack growth in brittle disordered materials (i) exhibits an intermittent dynamics with scale free distributions in space, time and energy and (ii) leads to rough fracture surfaces with scale invariant morphological features. As was exposed in Section \ref{Sec:5}, these two aspects can be qualitatively reproduced in simple numerical models such as RFM. However, these models rely on important simplifications which make quantitative comparisons with experiments difficult. 

We turn now to another approach - pioneered by Gao and Rice (1989) \cite{Gao89_jam} and later extended by Bower and Ortiz (1991) \cite{Bower91_jmps}, Schmittbuhl \etal (1995) \cite{Schmittbuhl95_prl}, Larralde and Ball (1995) \cite{Larralde95_epl}, Ramanathan \etal (1997) \cite{Ramanathan97_prl} and Lazarus (2003) \cite{Lazarus03_ijf} which consists in using LEFM to describe crack growth in an elastic isotropic material in presence of deterministic or random obstacles. As we will see below, to first order, the equation of motion and the equation of path are decoupled and can be solved independently. 

\begin{figure}
\begin{center}
\includegraphics[width=0.7\textwidth]{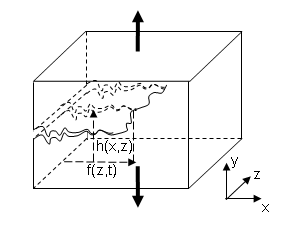}
\caption{Sketch and notation of a crack front propagating in a 3D material.}
\label{Sec:6:fig1}
\end{center}
\end{figure}

\subsection{Equation of motion}\label{Sec:6.1}

Let us consider the situation depicted in Figure \ref{Sec:6:fig1} of a crack front that propagates within a 3D elastic solid. Provided that the motion is slow enough, the local velocity of a point ${\bf M}\left(z,x=f(z,t),y=h(z,x=f(z,t))\right)$ is proportional to $G({\bf M}) - \Gamma({\bf M})$ (Equation \ref{Sec:2:equ5}). Obstacles and inhomogeneities in the material structure are then captured by introducing a fluctuating component into the fracture energy: $\Gamma({\bf M})=\Gamma^0\left(1+\eta({\bf M}) \right)$ where $\eta({\bf M})$ denotes the fluctuating part of fracture energy. This induces distortions of the front, which in turn generates perturbation in $G(\bf{M})$. One can then use the three-dimensional weight-function theory derived by Rice\cite{Rice85_jam} to relate $G(\bf{M})$ to $\{f,h\}$. To first order, $G(\bf{M})$ depends on the in-plane distortions $f(z,t)$ only and the resulting equation of motion reads \cite{Schmittbuhl95_prl,Ramanathan97_prl}:

\begin{equation}
\frac{1}{\mu} \frac{\partial f}{\partial t} = F+\frac{G^0}{2\pi}\int\frac{f(z)-f(z')}{(z-z')^2} \ud z'+\Gamma^0\eta(z,f(z,t)),
\label{Sec:6:equ1}
\end{equation}

\noindent where $F=G^0-\Gamma^0$ and $G^0$ denotes the reference mechanical energy release which would result from the same loading with a straight front at the same mean position. As it, $G^0$ depends on the macroscopic geometry and loading conditions. It is worth to mention that this equation of motion was recently shown by Dalmas and co-workers \cite{Dalmas09_jmps} to reproduce quantitatively experiments which consists in making a planar crack propagate along a weak interface with simple mono-dimensional regular patterns in a thin film stack (single band or periodic bands drawn parallel to the direction of crack propagation).  

\begin{figure}
\includegraphics[height=0.4\textwidth]{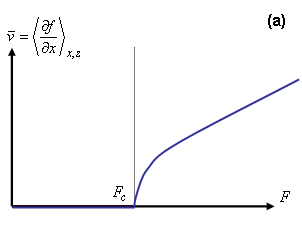}
\includegraphics[height=0.4\textwidth]{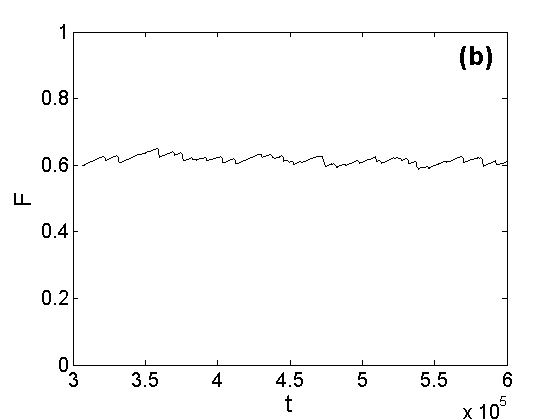}
\caption{(a) Pinning/depinning transition predicted by Equation \ref{Sec:6:equ1}. The control parameter in this system is the force $F$ defined as the difference between the stress intensity factor $K_I^0$ that would result from the same loading with a straight crack front at the same mean position and the mean value of the local toughness $K^0_c$. The order parameter is the mean value of the velocity $\overline{v}=\left\langle \partial f /\partial x \right\rangle_{z,t}$ averaged over all $z$ and $t$. (b) Time evolution of $F(t)$ in displacement imposed loading conditions (Obtained by solving numerically Eq. \ref{Sec:6:equ1} with $\mu=G^0=\Gamma^0=1$ and $F(t,\langle f \rangle)$ given by Eq. \ref{Sec:6:equ2} with $c=10^{-5}$ and $k=10^{-3}$).}
\label{Sec:6:fig2}
\end{figure}

In the case of heterogeneous materials, $\eta({\bf M})$ is chosen to be random, with short range spatial correlation, zero mean and constant variance. Let us first consider the situation of constant remote stress loading conditions, i.e. constant $F$ - This assumption will be released later in this section. As first noticed by Schmittbuhl \etal (1995) \cite{Schmittbuhl95_prl}, the front motion described by Equation \ref{Sec:6:equ1} exhibits a so-called depinning transition controlled by the "force" $F=G^0-\Gamma^0$ and its position with respect to a critical value $F_c$ (see Figure \ref{Sec:6:fig2}a).
\begin{itemize}
\item When $F < F_c$ the crack front is pinned by disorder and does not propagate.
\item When $F \gg F_c$ the crack front moves with a mean velocity $\bar{v}=\langle \partial f /\partial t\rangle_{z,t}$ proportional to $F$: $\bar{v}/\mu=G^0 - \Gamma^0$. One recovers the equation of motion expected by LEFM (Equation \ref{Sec:2:equ5}).
\item When $F = F_c$ a {\em critical} state is reached.
\end{itemize}

The consequences of this criticality will be extensively commented later in this section. But it is now time to discuss more explicitly the form taken by $G^0$ and its role in the intermittent crackling dynamics discussed in Section \ref{Sec:3}. In this respect, it has been suggested, in ref. \cite{Bonamy08_prl}, to consider the case of a crack growing stably in a material remotely loaded by imposing a constant displacement rate. This situation is the one encountered as e.g. in earthquakes problems where a fault is loaded because of the slow continental drift, or in the interfacial experiments described in Section \ref{Sec:3.3} where a crack front is made propagate along the weak heterogeneous interface between two Plexiglas block by lowering the bottom part at constant velocity. In this situation, $G^0$ is not constant anymore, but:
\begin{itemize}
\item increases with time since the remote loading increases with time.
\item decreases with the mean crack length $\langle f \rangle (t)=\langle f(z,t) \rangle_{z,t}$ since the material compliance decreases with $\langle f \rangle(t)$.
\end{itemize}
\noindent As a result, provided that the mean crack growth velocity is slow enough and the mean crack front is large enough, one can write \cite{Bonamy08_prl}:

\begin{equation}
F(t,\langle f \rangle)=ct-k \langle f \rangle (t),
\label{Sec:6:equ2}
\end{equation}

\noindent where $c$ and $k$ are constants depending on the precise geometry and the loading conditions (see e.g. \cite{Bonamy08_prl} for their value in the case of the interfacial experiment described in Section \ref{Sec:3.3}). The crack motion can then be decomposed as follows (Figure \ref{Sec:6:fig2}b): When $F \leq F_c$, the front remains pinned and the effective force increases with time. As soon as $F \geq F_c$, the front propagates, $\langle f \rangle(t)$ increases, and, as a consequence, $F$ is reduced. 
This retro-action process keeps the system attracted to the critical state, as in models of self-organized criticality \cite{Dickman00_bjp}. For high values of $c$, the front moves smoothly with mean velocity $\langle v \rangle=c/k$. For for low values of $c$, stick-slip motion is observed and the crack propagates through distinct avalanches between two successive pinned configurations. In the limit $c \rightarrow 0$ and $k \rightarrow 0$, criticality is reached and one expects to observe {\em universal} scaling features:
\begin{itemize}
\item independent of the microscopic and macroscopic details of the system, i.e. the considered material and the loading conditions
\item identical to those observed in other systems belonging to the same universality class, as e.g. interface motion in disordered magnets \cite{Urbach95_prl,Durin00_prl} and wetting of rough substrates \cite{Joanny84_jcp,Ertas94_pre}.
\end{itemize}

These scaling features can be predicted using theoretical tools issued from Statistical Physics like Functional Renormalization Methods (FRG) or numerical simulations. In particular, for $F$ slightly above $F_c$, the front $f(z)$ is self-affine and exhibits Family-Viseck dynamic scaling form up to a correlation length $\xi$:

\begin{equation}
\begin{array} {l}
\langle (f(z+\Delta z,t+\Delta t) - f(z,t))^2 \rangle^{1/2} \propto \Delta t^{\zeta_H/\kappa}g(\Delta z/\Delta t^{1/\kappa}) \\
\\
$where$\quad	g(u) \propto \left\{
\begin{array}{l l}
1 & $if u$ \ll c  \\
u^{\zeta_H} & $if u$ \gg c
\end{array}
\right.
\end{array}
\label{Sec:6:equ3}
\end{equation}

\noindent where $\zeta_H$ and $\kappa$ refer to the roughness exponent and the dynamic exponent respectively. It is worth to recall that $\zeta_H$ is associated to the in-plane roughness of the crack front and therefore is different from $\zeta$ defined in section \ref{Sec:4} to characterize the out-of-plane roughness. Above $\xi$, the front is no longer self-affine and exhibits logarithmic correlation.

The exponents $\zeta_H$ and $\kappa$ were estimated using FRG methods. They were found to be between $\{\zeta_H = 1/3,\kappa = 7/9\}$ \cite{Ertas94_pre} (first order) and $\{\zeta_H = 0.47,\kappa = 0.66\}$ \cite{Chauve01_prl} (second order). More recently, they were precisely evaluated by Rosso and Krauth \cite{Rosso02_pre} and Duemmer and Krauth \cite{Duemmer07_jsm} using numerical simulations: 

\begin{equation}
\zeta_H = 0.385 \pm 0.005, \quad \kappa = 0.770 \pm 0.005 
\label{Sec:6:equ4}
\end{equation} 

\noindent The correlation $\xi$ is set by the distance between the force $F$ and its critical value $F_c$. In situations where $F$ is kept constant, $\xi$ diverges at $F_c$ as:

\begin{equation}
\xi \propto (F-F_c)^{-\nu} \quad \mathrm{with} \quad \nu=\frac{1}{1-\zeta_H} = 1.625 \pm 0.010,
\label{Sec:6:equ5}
\end{equation} 

\noindent where the relation between $\nu$ and $\zeta_H$ comes from a specific symmetry of Equation \ref{Sec:6:equ1}, referred to as the statistical tilt symmetry \cite{Nattermann92_jpii}. In situations where $F$ varies with time and crack length according to equation \ref{Sec:6:equ2}, $\xi$ is set by the  "strength" of the feedback term. To be more precise, $\xi$ is defined \cite{Durin00_prl} as the length for which the feedback term $-k\langle f \rangle (t)$ in Equation \ref{Sec:6:equ2} balances the elastic term $(G^0/2\pi)\int(f(z)-f(z'))/(z-z')^2 \ud z'$ in Equation \ref{Sec:6:equ1}. This leads to $k f \xi/L \approx  G^0 f / \xi$ where $L$ refers to the specimen width. Hence, equation \ref{Sec:6:equ5} should be replaced by \cite{Durin00_prl}:

\begin{equation}
\xi \propto k^{-\nu_k} \quad \mathrm{with} \quad \nu_k=1/2
\label{Sec:6:equ6}
\end{equation} 

\begin{figure}
\begin{center}
\includegraphics[width=0.7\textwidth]{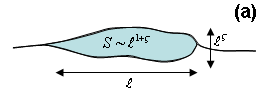}
\includegraphics[width=0.49\textwidth]{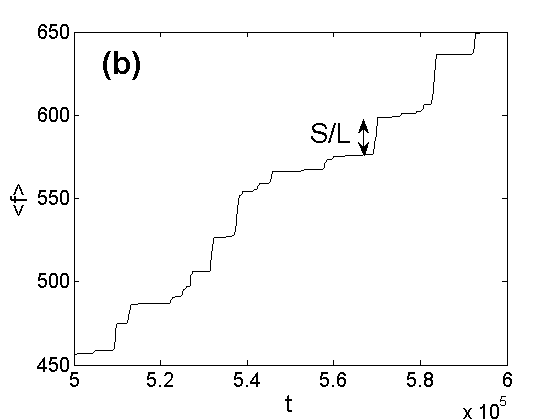}
\includegraphics[width=0.49\textwidth]{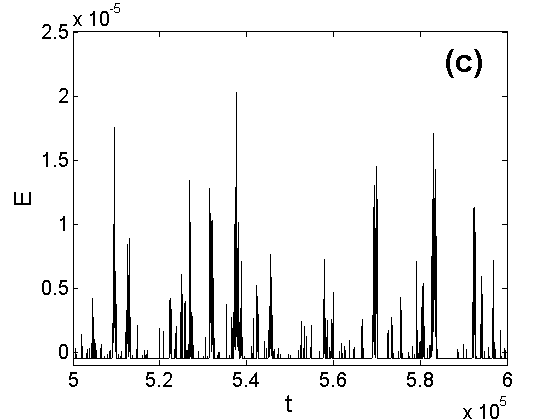}
\caption{(a) Sketch of the front propagation between two pinned configuration. The avalanche area $A$ scales with the size $\ell$ of the depinned zone as $A\propto \ell^{1+\zeta_H}$. (b) Typical evolution of the mean crack length. The crack progresses through jumps of size $S$ proportional to $A$. (c) Typical evolution of the energy released $E$. The energy released during an avalanche is also proportional to $S$.  Note the similarity between this curve and the one presented in figure \ref{Sec:3:fig1}a}
\label{Sec:6:fig3}
\end{center}
\end{figure}

The front propagation occurs through avalanches between two successive pinned configurations (Figure \ref{Sec:6:fig3}a). An avalanche of size $S$ translates into an increment $S/L$ for the mean crack length (Figure \ref{Sec:6:fig3}b). Equation \ref{Sec:6:equ2} ensures also that the mechanical energy $E$ released during an avalanche is also proportional to $S$ (Figure \ref{Sec:6:fig3}c). As a result, the energy signal is very similar to that observed in figure \ref{Sec:3:fig1}a.

The avalanche size $S$ is then shown \cite{Ertas94_pre} to obey power-law distribution: 

\begin{equation}
P(S) \propto S^{-\tau}f(S/S_0)
\quad \mathrm{with} \quad \tau=2 - \frac{1+2\nu\zeta_H}{\nu(1+\zeta_H)}= 1.280 \pm 0.010,
\label{Sec:6:equ7}
\end{equation} 

\noindent where the cut-off scales as $S_0\propto \xi^{1+\zeta_H}$, i.e., after having used Equation \ref{Sec:6:equ6}:

\begin{equation}
S_0 \propto k^{-1/\sigma}
\quad \mathrm{with} \quad \sigma= \frac{2}{1+\zeta_H} = 1.445\pm 0.005
\label{Sec:6:equ8}
\end{equation} 

\noindent The avalanche duration $T$ scales with the avalanche size $\ell$ as $T \propto \ell^{\kappa}$. Since the avalanche size $S$ scales as $\ell^{1+\zeta_H}$, one gets:

\begin{equation}
T \propto S^{a}
\quad \mathrm{with} \quad a= \frac{\kappa}{1+\zeta_H}=0.555\pm 0.005
\label{Sec:6:equ9}
\end{equation} 

\noindent And, after having changed the variables accordingly in equation \ref{Sec:6:equ7}, one shows that $T$ is distributed as: 

\begin{equation}
P(T) \propto T^{-\alpha}f(T/T_0) \quad \mathrm{with} \quad \alpha =1 + \frac{\nu-1}{\kappa\nu}  = 1.500 \pm 0.010,
\label{Sec:6:equ10}
\end{equation} 

\noindent where the cut-off scales as $T_0\propto \xi^{\kappa}$, which implies:

\begin{equation}
T_0 \propto k^{-1/\Delta} \quad \mathrm{with} \quad \Delta=1/\kappa\nu_k = 1.300 \pm 0.010 
\label{Sec:6:equ11}
\end{equation} 

\begin{figure}
\begin{center}
\includegraphics[width=0.8\columnwidth]{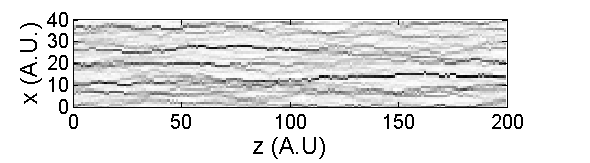}
\includegraphics[width=0.8\columnwidth]{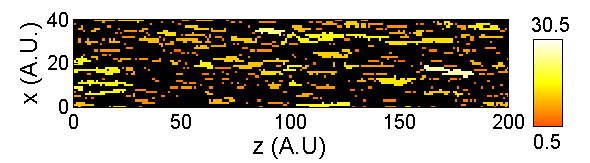}
\includegraphics[width=0.48\columnwidth]{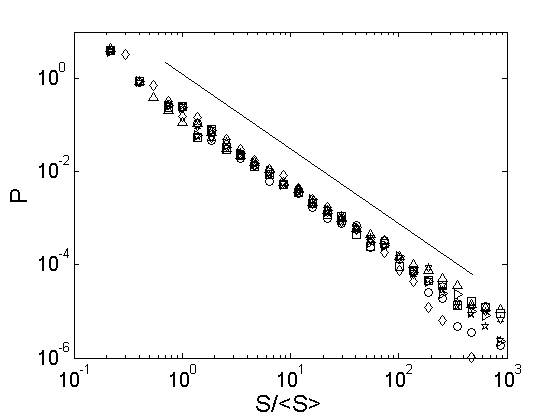}
\caption{(a) Typical gray scale map of the waiting time matrix $w(x,y)$ obtained from the solution of equations \ref{Sec:6:equ1} and \ref{Sec:6:equ2}. (b) Spatial distribution of clusters corresponding to velocities four times larger than the mean one. The clusters duration is given by the clusters color according to the colorscale given in inset. The distribution of the cluster size $S$ are plotted in (c). The various symbols corresponds to various values of $c$, $k$ and cluster thresholds. The straight lines correspond to $P(S_w)\propto S_w^{-\tau_w}$ with $\tau_w=1.7$ (From \cite{Bonamy08_prl}).}  
\label{Sec:6:fig4}
\end{center}
\end{figure}

Presently, these scaling predictions on the {\em global} dynamics of the crack front were not confronted directly to experiments. Work in progress \cite{Santucci09_prep} seems to indicate that the size and duration of the mean front jumps observed in the interfacial experiment presented in section \ref{Sec:3.3} are distributed according to equations \ref{Sec:6:equ7} and \ref{Sec:6:equ11}, respectively. On the other hand, the {\em local} dynamics of the stochastic description presented here was directly confronted to the interfacial experiment presented in section \ref{Sec:3.3} \cite{Bonamy06_prl}. The equation \ref{Sec:6:equ1} with $F$ given by equation \ref{Sec:6:equ2} was solved for various values of parameters and the results were analyzed using the procedure described in section \ref{Sec:3.3}: The map of waiting times (figure \ref{Sec:6:fig4}a), i.e. the time spent by the crack front in each point $(z,x)$ of the recorded region was computed, as in the experiments presented in section \ref{Sec:4.3}. Then, the avalanches were defined by thresholding these maps of waiting time, also as in section \ref{Sec:4.3} (figure \ref{Sec:6:fig4}b). The statistics of the obtained avalanches was finally computed and found to be the same as that observed in the interfacial experiment, with in particular a  power-law distribution for avalanches area characterized by an exponent $\tau_w \simeq 1.7$ (figure \ref{Sec:6:fig4}c). Presently, the relation between $\tau_w$ and the standard exponents defined in equations \ref{Sec:6:equ4} to \ref{Sec:6:equ11}  remains unknown.

It is finally worth to discuss the roughness exponent $\zeta_H \simeq 0.385$ expected for the in-plane crack front (equation \ref{Sec:6:equ4}). This value is significantly smaller than the value $\zeta_H \simeq 0.6$ reported in interfacial experiments \cite{Schmittbuhl97_prl,Delaplace99_pre}. However, recent experimental results \cite{Santucci09_submitted} shed light on the apparent disagreement between the two and report at large scales a roughness exponent $\zeta_H \simeq 0.35$ compatible with theoretical predictions of equation \ref{Sec:6:equ4}.

\subsection{Path equation} \label{Sec:6.2}  

We turn now to out-of-plane crack roughness. The universal morphological scaling features observed both in experiments and in lattice simulations reported in sections \ref{Sec:3} and \ref{Sec:4.2}, respectively, have lead to intense theoretical development over the last 20 years. We will focus here on the stochastic descriptions of crack path in 3D elastic disordered materials, in the spirit of what was done in the preceding section about motion equation, returning in section \ref{Sec:6.3} to a brief discussion of alternative interpretations of the morphological scaling features.

Let us consider again the situation depicted in figure \ref{Sec:6:fig1} where a crack front propagates within a 3D isotropic elastic solid remotely loaded in mode I. Provided that the motion is slow enough, PLS imposes that the path chosen by the crack is the one for which the net mode II stress intensity factor vanishes at each point M of the front at each time $t$ (section \ref{Sec:2.3}): $K_{II}({\bf M}(t))=0$. The tough part of the problem is to relate $K_{II}({\bf M}(t))$ to the crack roughness $f(z,t)$ and $h(z,x)$ (figure \ref{Sec:6:fig1}). This has been made possible thanks to the work of Larralde and Ball (1995) \cite{Larralde95_epl} later refined by Movchan \etal (1998) \cite{Movchan98_ijss} which shows that, to first order, $K_{II}({\bf M}(t))$ only depends on $h$. It can then be written as the sum of two contributions: A first contribution arises from the coupling of the singular mode I component $K_I^0$ of the stress field of the unperturbed crack with the position along the crack edge and a second contribution comes from the coupling between the slope of the crack surface and the non singular $T^0$ normal stress in the direction of crack propagation. Both contributions can be computed analytically (see \cite{Movchan98_ijss} for the complete form), but the second contribution was shown to be negligible compared to the first one when one considers roughness length-scales small with respect to the system size \cite{Larralde95_epl}. To the remaining contribution, a stochastic 3D term is added to capture the heterogeneous nature of the material, and a constant term $K^0_{II}$ is added to take into account the inherent misalignment in any loading system. Introducing all these terms in the PLS $K_{II}({\bf M}(t))=0$ leads to the following path equation \cite{Bonamy06_prl}:

\begin{equation}
\frac{\partial h}{\partial x}=\frac{K^0_{II}}{K^0_{I}} + \frac{1}{\pi}\frac{2-3\nu}{2(2-\nu)} \int_{-\infty}^\infty \frac{h(z)-h(z')}{(z-z')^2} \ud z'+\eta(z,x,h(z,x))
\label{Sec:6:equ12}
\end{equation}

\noindent At this point, it is worth to recall that this path equation was derived within elastostatic and, as such, cannot work for dynamic fracture when the crack speed exceeds a few percent of the Rayleigh wave speed.

The form taken by this equation is very similar to that of Eq. \ref{Sec:6:equ1} which describes the in-plane motion of the crack front. Two main differences are however worth to be discussed:
\begin{itemize}
\item[(i)] The absence of explicit time $t$ in equation \ref{Sec:6:equ12}. This comes from the fact that to first order, $K_{II}$ only depends on the out-of-plane roughness $h(z,x)$. This yields a decoupling between this path equation and equation \ref{Sec:6:equ1} describing the crack dynamics. It implies that the scaling properties of fracture surfaces will not depend on the crack velocity $v$ or crack loading $G^0$, contrary to those of the in-plane projection of the crack front. 
\item[(ii)] The 3D nature of the stochastic term $\eta(z,x,h(z,x))$. 
\end{itemize}
\noindent This second point makes equation \ref{Sec:6:equ12} extremely difficult - if not impossible - to solve. Two limit cases can however be considered. 

The first limit consists \cite{Larralde95_epl,Ramanathan97_prl} in considering very smooth fracture surfaces so that the 3D stochastic term $\eta(z,x,y=h(z,x))$ reduces to an effective "thermal" term $\eta(z,x)$. Fracture surfaces are then predicted to exhibit logarithmic scaling. This prediction seems in apparent disagreement with the experimental observations reported in section \ref{Sec:4}. However, recent observations performed by Dalmas and coworkers (2008) \cite{Dalmas08_prl} on nanoscale phase separated glasses reveals logarithmic roughness at large length scales.

The second approximation consists \cite{Bonamy06_prl} in writing $\eta(z,x,y=h(z,x))$ as the sum of two 2D terms: $\eta(z,x,y=h(z,x))=\eta(z,x)+\eta(z,y=h(z,x))$. The morphology of the fracture surface $h(x,z)$ is then given by the motion of the elastic string $h(z)$ that "creeps" - the $x$ coordinate playing the role of time - within a random potential $\eta(z,h(z,x))$ due to the "thermal" fluctuations $\eta(z,x)$. The fracture surface $h(z,x)$ obeys then to Family-Viseck anisotropic self-affine scaling (equation \ref{Sec:4:equ5}) with $\{\zeta_{\perp} \simeq 0.385, \zeta_{||} \simeq 0.5\}$ in perfect agreement with observations reported at large scales, as e.g. in sandstone \cite{Boffa98_epjap,Boffa00_pa,Ponson07_pre}, in artificial rocks \cite{Bouchbinder06_prl} and granular packing made of sintered glass beads \cite{Ponson06_prl2,Bonamy06_prl,Ponson07_AnPhys} (see also section \ref{Sec:4.3}).

Let us finally mention that similar stochastic approach has been developed recently by Katsav \etal \cite{Katzav07_epl} in simpler 2D elastic materials. In this case, the path equation can be solved analytically and leads to apparent self-affine scaling features at small scales, with $\zeta=0.5$, and departure from self-affinity at large scales. However, these predictions seem in contradiction with experimental observations \cite{Kertesz93_fractals,Salminen03_epjb,Santucci07_pre} which report a unique $\zeta \simeq 0.7$ self-affine scaling regime in quasi-two-dimensional materials.

\subsection{Alternative interpretations for self-affine crack roughness}\label{Sec:6.3}

The stochastic theory of crack growth in elastic brittle disordered materials described in section \ref{Sec:6.2} seems to reproduce quite well the morphological scaling properties of cracks for length-scale above the FPZ size. On the other hand, it cannot describe observations performed below the FPZ size. In particular, in essence, it cannot reproduce the small-scale anisotropic Family-Viseck scaling characterized by a roughness exponent $\zeta \simeq 0.8$ reported in a wide range of materials and discussed in section \ref{Sec:4}. In this respect, we will present now some alternative approaches that were developed to account for the morphological scaling features observed in fractography experiments as well as in RFM simulations.

Hansen \etal (1991) \cite{Hansen91_prl} have studied the morphology of crack lines (2D case) or crack surfaces (3D) obtained after the breakdown of RFM with perfectly {\em plastic} fuses, i.e. fuses that act as unit Ohmic resistors up to a threshold voltage and that carry a constant current above. They have shown that the problem can be mapped to that of a directed polymer (2D case) or a minimum energy surface (3D case) in random potential. As a result, crack path can be mapped to a KPZ equation \cite{Kardar86_prl} and crack lines/surfaces are expected to display self-affine morphological isotropic scaling features with a universal roughness exponent equal to $\zeta=2/3$ in 2D, and $\zeta \simeq 0.42$ in 3D. They conjectured that these results, demonstrated rigorously on elastic-perfect-plastic model, can also be relevant for brittle fracture. Indeed, these predictions are in close (but not perfect) agreement with the $\zeta \simeq 0.71$ (2D) and $\zeta \simeq 0.42$ (3D) obtained from numerical simulations of brittle RFM and presented in section \ref{Sec:4.2}. However, the anomalous scaling is not recovered. Hansen \etal's predictions are also very close to the $\zeta \simeq 0.7$ observed experimentally in 2D sheet of paper \cite{Kertesz93_fractals,Salminen03_epjb,Santucci07_pre} and the $\zeta \simeq 0.4-0.5$  observed in sandstone \cite{Boffa98_epjap,Boffa00_pa,Ponson07_pre} or glassy ceramics made of sintered glass beads \cite{Ponson06_prl2,Bonamy06_prl,Ponson07_AnPhys}. 

Bouchaud \etal (1993) \cite{Bouchaud93_prl} proposed to model the fracture surface as the trace left by a line moving through a 3D disordered landscape  - the dynamics of which is described through a phenomenological nonlinear Langevin equation derived initially by Ertas and Kardar \cite{Ertas92_prl} to describe the motion of vortex lines in superconductors. 
The resulting fracture surfaces are then predicted to display anisotropic Family-Viseck scaling with $\{\zeta_\perp\simeq 0.75$, $\zeta_{||} \simeq 0.5\}$.
This first work was then extended by Daguier \etal (1997) \cite{Daguier97_prl} who showed that, depending on the crack velocity $v$, two distinct self-affine regimes can be expected: A large length-scale regime with $\{\zeta_\perp\simeq 0.75$, $\zeta_{||} \simeq 0.5\}$, and a small length-scale regime with $\{\zeta^c_{\perp}\simeq 0.5$, $\zeta^c_{||} \simeq 0.4\}$. The crossover $\xi$ between the two is expected \cite{Daguier97_prl,Bouchaud97_jpcm} to diverge with $v$ as $\xi \propto v^{-1/(\zeta^c_{\perp}/\zeta^c_{||}-\zeta^c_{\perp})}$. The exponents values obtained at large scale/large velocities are in excellent agreement with those observed in many brittle materials at scales smaller than the FPZ size (see section \ref{Sec:4.3}). However, the existence of a second small scale regime all the more important than crack velocity is large is incompatible with experiments.   

More recently, Hansen, Schmittbuhl and Batrouni (2003) \cite{Hansen03_prl,Schmittbuhl03_prl} suggested that the universal scaling properties of fracture surfaces are due to the fracture propagation being a damage coalescence process described by a stress-weighted percolation phenomenon in a self-generated quadratic damage gradient. 

Then, assuming that the damage profile is proportional to the system size $L$, the rms roughness $w$ of the crack line/surface is expected to scale as:

\begin{equation}
w \propto L^\zeta \quad with \quad \zeta=\frac{2\nu_{perc}}{1-2\nu_{perc}},
\label{Sec:6:equ13}
\end{equation}

\noindent where $\nu_{perc}$ is the exponent describing the divergence of the correlation length $\xi_{perc} \propto (p-p_c)^{-\nu_{perc}}$ as the density of broken bonds $p$ is reaching the critical value $p_c$ in standard percolation problems (see e.g. \cite{Stauffer94_book} for an introduction to percolation theory).

In percolation theory, the exponent $\nu_{perc}$ depends only on the system dimension and  dimensionality. In standard {\em scalar} percolation, $\nu_{perc}=4/3$ in 2D, and $\nu_{perc}=0.88$ in 3D \cite{Stauffer94_book}. This leads to $\zeta=8/11$ in 2D, and $\zeta \simeq 0.64$ in 3D. The first value is compatible to the $\zeta \simeq 0.7$ measured in 2D brittle RFM (see section \ref{Sec:4.2} or references \cite{Zapperi05_pre,Nukala07_pre}), while the second one is significantly higher than the $\zeta \simeq 0.42$ observed in 3D brittle RFM (see section \ref{Sec:4.2} or references \cite{Raisanen98_prl,Raisanen98_prb,Nukala06_pre}). In tensorial {\em elastic} percolation, $\nu_{perc}=2$ in 3D, which leads to $\zeta=4/5$. The value of the roughness exponent is in excellent agreement with $\zeta \simeq 0.8$ observed in many brittle materials observed at scales smaller than the FPZ size (see section \ref{Sec:3}), but the Family-Viseck anisotropic scaling is not captured.

This interpretation yields recent controversies. First, the analysis of large-scale RFM simulations \cite{Nukala04_jsm} suggests that, contrary to what was assumed, (i) damage profile is not quadratic and (ii) the width of the damage profile is not linear with the system size. Second, it was mentioned \cite{Alava04_prl} that the exponent $\zeta$ in equation \ref{Sec:6:equ13} cannnot be intepreted as a "standard" roughness exponent since, strickly speaking, the front is not self-affine. Indeed, in such a model, the fracture perimeter displays substantial overhangs the size of which is comparable with their width. To be more precise, self-similarity (i.e. isotropic scaling) is expected for length scales smaller than $w$, and a trivially flat front beyond. In response, Hansen, Schmittbuhl and Batrouni \cite{Schmittbuhl04_prl} pointed out that experimental fracture fronts are also found to display significant overhangs while remaining self-affine at large scales. They argue that the non-isotropic scaling given by equation \ref{Sec:6:equ13} is sufficient to define self-affine surfaces.    

\begin{landscape}
\begin{table}
\begin{footnotesize}
\begin{center}
\begin{tabular}{|l|cccc|}
 \cline{2-5}
 \multicolumn{1}{l|}{}   &  & & & \\           
 \multicolumn{1}{l|}{}   & Energy $^{(i)}$ & Size $^{(ii)}$ & Silent time $^{(iii)}$ & Roughness $^{(iv)}$ \\            
 \multicolumn{1}{l|}{}   &  & & & \\           
 \hline
 & & & & \\
Earthquakes \cite{Utsu99_pag,Benzion08_rg} & $\beta \simeq 1.7$ & $\tau \simeq 2$ & $\alpha \simeq 1$ & - \\
 & & & & \\
\hline
 & & & & \\
Lab experiments, 3D & non-universal $\beta_{AE}$ & - & non-universal $\alpha$ & small scale: $\zeta_{\perp} \simeq 0.75$, $\zeta_{||} \simeq 0.6$ \cite{Ponson06_prl,Ponson06_prb,Ponson06_ijf} \\
                    &                            &   &                        & large scale: $\zeta_{\perp} \simeq 0.4$, $\zeta_{||} \simeq 0.5$ \cite{Bonamy06_prl,Ponson06_prl2,Boffa98_epjap}, or logarithmic \cite{Dalmas08_prl}\\
 & & & & \\
Lab experiments, interfacial & - & $\tau_W \simeq 1.7$ \cite{Grob09_pag} &  $\alpha \simeq 1$ \cite{Grob09_pag} & small scale: $\zeta_H \simeq 0.6$ \cite{Schmittbuhl97_prl,Delaplace99_pre} \\
                             &   &                     &                     & large scale: $\zeta_H \simeq 0.35$ \cite{Santucci09_submitted} \\
 & & & & \\
Lab experiments, 2D & $\beta_{AE} \simeq 1$ \cite{Salminen02_prl} & $\tau \simeq 1.5$ \cite{Salminen02_prl} &  non-universal $\alpha$ & $\zeta \simeq 0.7$ \cite{Kertesz93_fractals,Salminen03_epjb,Santucci07_pre} \\
 & & & & \\
\hline
 & & & & \\
RFM simulations, 3D & - & $\tau=2.5$ \cite{Zapperi05_pa} & - & $\zeta \simeq 0.42$ \cite{Raisanen98_prl,Raisanen98_prb,Nukala06_pre} \\ 
 & & & & \\
RFM simulations, 2D & - & non universal \cite{Zapperi05_pre} & - & $\zeta \simeq 0.71$ \cite{Zapperi05_pre,Nukala07_pre} \\ 
 & & & & \\
\hline
 & & & & \\
stochastic LEFM, 3D & $\beta \simeq 1.280$ & $\tau\simeq 1.280$, $\tau_W \simeq 1.7$ \cite{Bonamy08_prl} & - & $\zeta_{\perp} \simeq 0.385,\zeta_{||} \simeq 0.5$ \cite{Bonamy06_prl} or logarithimic \cite{Larralde95_epl,Ramanathan97_prl} \\ 
 & & & & \\
stochastic LEFM, interfacial & $\beta \simeq 1.280$ & $\tau\simeq 1.280$, $\tau_W \simeq 1.7$ \cite{Bonamy08_prl} & - & $\zeta_H \simeq 0.385$ at small $v$ \cite{Schmittbuhl95_prl,Ramanathan97_prl}\\ 
                             &                    &                                         &   & logarithmic at high $v$ \\ 
 & & & & \\
stochastic LEFM, 2D & no powerlaw & no powerlaw & no powerlaw & small scale: $\zeta = 0.5$ \\ 
                    &             &             &             & large scale: not self-affine \cite{Katzav07_epl} \\ 
 & & & & \\
\hline
\end{tabular}
\end{center}
\begin{tabular}{rp{0.9\linewidth}}
$^{(i)}$   & The exponent $\beta$ characterizes the distribution of released energy during fracture events. The exponent $\beta_{AE}$ characterizes the energy distribution of AE triggered by these fracture events. The relation between the two remains unknown.\\  
$^{(ii)}$   & The exponent $\tau$ characterizes the distribution of rupture area associated with fracture events. The exponent $\tau_{W}$ characterizes the distribution of quake events defined from the waiting time map (see section \ref{Sec:3.3} for details). The relation between the two remains unknown.\\  
$^{(iii)}$   & The exponent $\alpha$ characterizes the distribution of silent time between two successive fracture events.\\  
$^{(iv)}$   & The exponent $\zeta$ refers to the roughness exponent measured on self-affine crack profiles (2D) or isotropic self-affine fracture surfaces (3D). When fracture surfaces exhibit anisotropy in scaling, $\zeta_{||}$ and $\zeta_{\perp}$ refer to the roughness exponents measured parallel and perpendicular to the direction of crack propagation, respectively. The exponent $\zeta_H$ refers to the roughness exponent measured on crack fronts in interfacial geometries.\\  
\end{tabular}
\caption{Summary of the exponents characterizing the various scale-free distributions and morphological scaling features reported in fields observations, laboratory experiments, numerical simulations and stochastic models derived from LEFM.}
\label{Tab:1}
\end{footnotesize}
\end{table}
\end{landscape}

\section{Concluding discussion}\label{Sec:7}

Many experiments and fields observations have revealed that brittle failure in materials exhibit scale-invariant features (see Tab. \ref{Tab:1} for a summary). In particular:(i) Fracturing systems displays jerky {\em crackling} dynamics with random impulsive energy release, as suggested from the AE accompanying the failure of various materials and, at much larger scale, the seismic activity associated to earthquakes. The energy distribution of these discrete events forms a power-law that spans over many orders of magnitude, with no characteristic scale and (ii) roughness of cracks exhibits self-affine morphological features, characterized by roughness exponents. These observations are common to many brittle materials and can be reproduced qualitatively in the electrical breakdown of Random Fuse Network. On the other hand, they cannot be captured by standard LEFM continuum theory.

Some of these scale-free distributions and scale-invariant morphological features are universal. Others are not. The experiments and lattice simulations presented in sections \ref{Sec:3}, \ref{Sec:4} and \ref{Sec:5} allows to distinguish two cases:
\begin{itemize}
\item[(A)] Damage spreading processes within a brittle material preceding the initiation of a macroscopic crack. The associated micro-fracturing events release energy impulses which are power-law distributed. The associated exponent is non-universal, but depends on the considered materials, loading conditions environment parameters (see section \ref{Sec:3.2})... This transient damage spreading is also suggested \cite{Ponson06_ijf,Morel08_pre} to be responsible for the anomalous non-universal scaling exhibited in the initial transient roughening regime of fracture surfaces following crack initiation \cite{Lopez98_pre,Morel98_pre,Morel08_pre}.      
\item[(B)] Macroscopic crack growth within a brittle material. When this propagation is slow enough, it exhibits an intermittent crackling dynamics, with sudden jumps and energy release events the distributions of which form power-law with apparent universal exponents. Crack growth leads to rough fracture surfaces that display Family-Viseck universal scaling far enough from crack initiation.  
\end{itemize}
Quite surprisingly, seismicity associated with earthquakes seems to belong to the second case and exhibits quantitatively the same statistical scaling features as that observed in experiments of interfacial crack growth along weak disordered interfaces \cite{Grob09_pag}.

The (non-universal) scaling features associated with the microfracturing events observed experimentally in case (A) are reproduced qualitatively in lattice models such as the RFM presented in section \ref{Sec:4}. Nevertheless, what sets precisely the distribution and the value of scaling exponents in experiments remains largely unknown. On the other hand, the recent stochastic extensions of LEFM theory presented in section \ref{Sec:6} seems efficient to capture quantitatively the observations performed in case (B). In these stochastic LEFM descriptions, the onset of crack growth is found to be analogous to a critical transition between a stable phase where the crack remains pinned by the material heterogeneities and a moving phase where the mechanical energy available at the crack tip is sufficient to make the front propagate \cite{Schmittbuhl95_prl,Ramanathan97_prl}. While growing, the crack decreases its mechanical energy and gets pinned again. Provided that the growth is slow enough, this retro-action process keeps the system close to the critical point during the whole propagation, as for self-organized-critical systems \cite{Bak87_prl}. As seen in section \ref{Sec:6.1}, the resulting dynamics exhibits spatio-temporal intermittency characterized by universal statistical features that can be predicted theoretically using FRG methods and are compatible with experimental observations.

The analogy between the onset of crack propagation and a critical dynamic transition has numbers of other consequences. Roux, Charles \etal (2003,2004) \cite{Roux03_ejma,Charles04_jmps} made use of the universality manifested around the depinning onset to determine the distribution of effective macroscopic toughness in heterogeneous materials. Their main results are: (i) Universal scaling between toughness variance and specimen size, and (ii) the existence of a universal toughness distribution for large enough specimens. These predictions were shown \cite{Charles06_ijf} to reproduce fairly well the statistics of crack arrest lengths observed in indentation experiments performed in various brittle materials. More recently, Ponson \etal (2007,2009)
\cite{Ponson07_proc,Ponson08_condmat} investigate the role of a finite temperature in this stochastic LEFM description of crack growth and propose a creep law to relate crack velocity and stress intensity factor in sub-critical failure regime. This was shown recently to describes rather well experiments of paper peeling \cite{Koivisto07_prl} and subcritical crack growth in sandstone \cite{Ponson08_condmat}.

It should be emphasized that the stochastic descriptions presented in section \ref{Sec:6} were developed within elastostatic approximation. As such, they cannot account for the dynamic
stress transfers through acoustic waves occurring as a dynamically growing crack is interacting with the material disorder \cite{Ravichandar84c_ijf,Fineberg92_prb, Sharon01_nature,Bonamy03_prl,Bonamy05_ijf}. The recent availability of analytical solutions for weakly distorted cracks within full 3D elastodynamics framework \cite{Willis95_jmps,Willis97_jmps} suggests promising developments in this context in a close future (see \cite{Ramanathan97b_prl,Bouchaud02_jmps} for recent theoretical work in this context).

As presented in section \ref{Sec:6.2} this stochastic LEFM description seems to capture fairly well the self-affine scaling features exhibited by fracture surfaces. In particular, the apparent disagreements between its predictions and experimental observations presented in Bouchaud's 1997 review \cite{Bouchaud97_jpcm} are now uncovered \cite{Bonamy06_prl,Dalmas08_prl}: The ``classical" self-affine scaling
regime characterized by a roughness exponent $\zeta \simeq 0.8$ widely reported in the literature is in fact observed below the FPZ size, where in essence stochastic LEFM descriptions stop to be relevant. 

While the problem of crack propagation in disordered brittle materials has been widely addressed theoretically, the role of disorder on the formation of seed cracks remains far less studied (see \cite{Arndt01_prb,Kierfeld06_prl} for recent theoretical work in this topic). The fact that fracture surfaces observed at small scale exhibit a universal small scale $\zeta \simeq 0.8$ Family-Viseck scaling in very different materials with various
damage processes as, e.g., plastic deformation, crack blunting, ductile cavity growth or microcracking remains largely unexplained. It suggests that it may be possible to find a generic unified statistical description of damage spreading within disordered brittle materials within the FPZ, independent of the precise nature of the considered material. In this context, damage descriptions in term of universal stress-weighted percolation processes as proposed by Schmittbuhl and Hansen (2003) \cite{Schmittbuhl03_prl} appear promising. Their development, their implications in terms of damage dynamics, and their careful confrontation to experiments and/or simulations represent interesting challenges for future investigations.

\ack
I would like to acknowledge Cindy Rountree, Frederic Lechenault and Laurent Ponson for a critical reading of this review, and thank all my collaborators, namely Harold Auradou, Luc Barbier, Elisabeth Bouchaud, Fabrice Célarié, Stéphane Chapuliot, Davy Dalmas, Claudia Guerra, Jean-Pierre Hulin, Laurent Ponson, Stéphane Morel, Cindy Rountree, Stéphane Santucci and Julien Scheibert. Special thanks to Stéphane Santucci and Knut-Jorgen M{\aa}l{\o}y who allowed me to reproduce their figures. This research activity is partly financed through ANR project RUDYMAT
(Grant No. ANR-05-JCJC-0088), and through ``Triangle de la Physique" (Projet CODERUP No. 2007-46). Finally, I would like to thank Luc Barbier and Francois Daviaud for their constant support.

\section*{References}

\bibliographystyle{unsrt}
\bibliography{bibfracture}

\end{document}